\documentclass[12pt]{iopart}
\usepackage{epsf}
\newcommand{\be}{\begin{equation}}
\newcommand{\ee}{\end{equation}}
\newcommand{\bea}{\begin{eqnarray}}
\newcommand{\eea}{\end{eqnarray}}

\begin{document}

\title[One and Two-Electron Spectral Function Expressions...]{One and
Two-Electron Spectral Function Expressions in the Vicinity of
the Upper-Hubbard Bands Lower Limit}
\author{J M P Carmelo\dag,\,L M Martelo\dag\ddag and P D Sacramento\S}
\address{\dag GCEP-Center of Physics, University of Minho, Campus
Gualtar, P-4710-057 Braga, Portugal}
\address{\ddag Physics Department, Engineering Faculty of University
of Porto, P-4200-465 Porto, Portugal}
\address{\S Departamento de F\'{\i}sica and CFIF, Instituto
Superior T\'ecnico, P-1049-001 Lisboa, Portugal}

\date{30 July 2003}


\begin{abstract}
In this paper we derive general expressions for few-electron
spectral functions of the one-dimensional Hubbard model for values
of the excitation energy in the vicinity of the $M^{th}$
upper-Hubbard band lower limit. Here $M=1,2,...$ is the
rotated-electron double occupation, which vanishes for the ground
state and is a good quantum number for all values of the on-site
Coulomb repulsion $U$. Our studies rely on a combination of
symmetries of the model with a recent finite-energy holon and
spinon description of the quantum problem. We apply our general
scheme to the one-electron addition spectral function, dynamical
structure factor, and spin singlet Cooper pair addition spectral
function. Our results provide physically interesting information
about the finite-energy spectral properties of the many-electron
one-dimensional quantum liquid.
\end{abstract}


\maketitle
\section{INTRODUCTION}

Recent experimental studies of quasi-one-dimensional (1D)
materials observed unusual finite-energy/frequency spectral
properties, which are far from being well understood
\cite{Hussey,Menzel,Fuji02,Hasan,Ralph}. Several of these
experimental studies reveal the occurrence of charge-spin
separation in terms of independent holon and spinon excitation
modes \cite{Menzel,Hasan,Ralph,I,IIa,IIIb,spectral0,spectral}. For
values of the excitation energy larger than the transfer integrals
associated with electron hopping between the chains, the 1D
Hubbard model \cite{Lieb} is expected to provide a good
description of the physics of these materials \cite{Hasan,Ralph}.
Unfortunately, an accurate determination of the spectral
properties for finite values of the excitation energy and of the
Coulomb repulsion is until now still lacking. Most accurate
results correspond to the limit of infinite on-site Coulomb
repulsion \cite{Penc97} where the Bethe-ansatz \cite{Lieb} wave
function is easier to handle, yet the solution of the problem
remains complex in this limit.

In this paper we combine the holon, spinon, and pseudoparticle
accurate description recently introduced and studied in Refs.
\cite{I,IIa} and the pseudofermion representation very recently
introduced in Ref. \cite{IIIb} with symmetries of the 1D Hubbard
model to derive general expressions for few-electron spectral
functions in the vicinity of the lower limit of the upper Hubbard
bands. For finite values of the excitation energy and on-site
repulsion $U$ there are not many previous studies of few-electron
spectral function weight distributions. The interesting studies of
the one-electron spectral functions presented in Ref.
\cite{Penc97} refer to the limit of infinite $U$. There are
numerical studies of these functions for $U\approx 4t$
\cite{Senechal}. Moreover, the weight distribution of the real
part of the frequency dependent optical conductivity in the
vicinity of the optical pseudogap was previously studied in Refs.
\cite{optical,opticalthey}. The very recent results presented in
Ref. \cite{spectral} refer to the one-electron spectral-function
weight-distribution in the vicinity of singular branch lines. Some
of these lines cross the upper-Hubbard band lower-limit points
considered in this paper. However, our results are complementary
of those presented in Ref. \cite{spectral}, which were obtained by
use of the method of Ref. \cite{V}. Indeed, our
weight-distribution expressions are valid when one approaches
these points along all finite-weight directions except those
corresponding to the above-mention singular branch lines. The
values of the critical exponents which control the weight
distribution are different for these lines and for all the
remaining directions of the ($k,\,\omega$)-plane considered here.

We evaluate expressions for the critical exponents that control
the weight distribution of few-electron spectral functions for
excitation energies corresponding to the vicinity of the $M^{th}$
upper Hubbard band lower-limit points. (The method used in this
paper is different from that of Ref. \cite{V}, which was used in
Ref. \cite{spectral} in the study of the weight distribution in
the vicinity of the singular branch lines.) Our studies rely on a
symmetry which is specific for the model when defined in the
reduced Hilbert subspace associated with the ($k,\,\omega$)-plane
region in the vicinity of the above lower-limit points.
Few-electron spectral functions can include several upper Hubbard
bands, the $M^{th}$ band being spanned by excited states of
rotated electron double occupation $M>0$. The lower Hubbard band
corresponds to the spectral-weight distribution generated by
excited states of rotated-electron double occupation $M=0$. For
one-electron and two-electron spectral functions only the first
few upper bands have a significant amount of spectral weight
\cite{V}. Although our general expressions refer to the
lower-limit points of all upper Hubbard bands, we apply our method
only to the study of weight distributions corresponding to upper
bands with a significant amount of spectral weight. For the
one-electron addition spectral function and dynamical structure
factor (and the spin singlet Cooper pair addition spectral
function) only the $M=1$ first Hubbard band has (and both the
$M=1$ first and $M=2$ second upper Hubbard bands have) a
significant amount of weight.

Our motivation in studying the finite-energy one-electron spectral
function is that our method provides the exponents of the singular
spectral features beyond the branch lines considered in Ref.
\cite{spectral}. The latter lines are observed in quasi-1D
materials \cite{spectral0,spectral}. Thus, the study of other
finite-energy singular spectral features is of interest for the
further understanding of the unusual spectral properties observed
in these materials. Both for the dynamical structure factor and
other spectral functions we do not consider the low-energy
features because those can be investigated by standard
two-component conformal-field theory. Our aim in studying the
Cooper-pair spectral function is to clarify whether there are
singular spectral features at the upper-Hubbard bands lower limit.
Indeed, in the limit of small on-site repulsion and for electronic
densities very close to one, such singular features would appear
at low energy and could lead to a superconductivity instability in
a system of weakly coupled Hubbard chains, as further discussed in
later sections.

The paper is organized as follows: In Sec. II we summarize the 1D
Hubbard model. In Sec. III we introduce the holon-spinon
Hamiltonian and study the 1D Hubbard model spectrum in the
vicinity of the lower limit of the upper Hubbard bands. In Sec. IV
we use the holon and spinon conservation laws and other symmetries
of the model in the evaluation of general expressions for
few-electron spectral functions. Moreover, we apply our method to
the study of the one-electron addition spectral function,
dynamical structure factor, and singlet Cooper pair addition
spectral function. Finally, Sec. V contains the concluding
remarks.


\section{THE 1D HUBBARD MODEL}

In a chemical potential $\mu $ and magnetic field $H$ the 1D
Hubbard Hamiltonian can be written as,
\bea
& &\hat{H} = {\hat{H}}_{SO(4)} + \sum_{\alpha=c,\,s}\mu_{\alpha
}\,{\hat{S}}^{\alpha}_z \, ; \nonumber \\
& & {\hat{H}}_{SO(4)} =
-t\sum_{j =1}^{N_a}\sum_{\sigma =\uparrow
,\downarrow}[c_{j,\,\sigma}^{\dag}\,c_{j+1,\,\sigma} +  c^{\dag
}_{j+1,\,\sigma}\,c_{j,\,\sigma}]+ U\,\sum_{j
=1}^{N_a}[\hat{n}_{j,\,\uparrow}-1/2]\,[\hat{n}_{j,\,\downarrow}-1/2]
\, , \label{H} \nonumber \\
& &
\eea
where $c_{j,\,\sigma }^{\dagger }$ and $c_{j,\,\sigma}$ are the
spin $\sigma $ electron creation and annihilation operators at
site $j$, respectively. The operator $\hat{n}_{j,\,\sigma}=
c_{j,\,\sigma }^{\dagger }\,c_{j,\,\sigma }$ counts the number of
spin $\sigma$ electrons at real-space lattice site $j=1,...,N_a$.
The number of lattice sites $N_a$ is even and large and $N_a/2$ is
odd. We consider periodic boundary conditions. Moreover,
$\mu_c=2\mu$, $\mu_s=2\mu_0 H$, $\mu_0$ is the Bohr magneton, and
${\hat{S }}^c_z= -(1/2)[N_a-\hat{N}]$ and ${\hat{S }}^s_z=
-(1/2)[{\hat{N}}_{\uparrow}-{\hat{N}}_{\downarrow}]$ are the
diagonal generators of the $\eta$-spin and spin $SU(2)$ algebras
\cite{HL}, respectively. The operators ${\hat{N}}=\sum_{\sigma}
\hat{N}_{\sigma}$ and ${\hat{N}}_{\sigma}=\sum_{j}
\hat{n}_{j,\,\sigma}$ count the number of electrons and spin
$\sigma$ electrons, respectively. The Hamiltonian
$\hat{H}_{SO(4)}$ defined in Eq. (\ref{H}) commutes with the six
generators of the $\eta$-spin and spin algebras \cite{HL}. The
Bethe-ansatz solvability of the 1D Hubbard model is restricted to
the Hilbert subspace spanned by the lowest-weight states (LWSs) of
the $\eta$ spin and spin algebras, {\it i.e.} such that
$S^{\alpha}= -S^{\alpha}_z$ \cite{I} for $\alpha =c$ and $\alpha
=s$, respectively. Here $S_c$ (and $S_s$) denotes the $\eta$-spin
(and spin) value. The momentum operator reads,

\begin{equation}
\hat{P} = \sum_{\sigma=\uparrow ,\,\downarrow }\sum_{k}\,
\hat{N}_{\sigma} (k)\, k \, , \label{Popel}
\end{equation}
where the spin $\sigma $ momentum distribution operator is given
by $\hat{N}_{\sigma} (k) = c_{k,\,\sigma }^{\dagger
}\,c_{k,\,\sigma}$ and the operator $c_{k,\,\sigma}^{\dagger}$
(and $c_{k,\,\sigma}$) creates (and annihilates) a spin $\sigma $
electron at momentum $k$. The momentum operator (\ref{Popel})
commutes with the Hamiltonians of Eq. (\ref{H}).

There are $N_{\uparrow}$ spin-up electrons and $N_{\downarrow}$
spin-down electrons in the chain of $N_a$ sites, lattice constant
$a$, and length $L=[N_a\,a]$ associated with the model (\ref{H}).
Throughout this paper we use units of Planck constant one and of
lattice spacing $a=1$ and denote the electronic charge by $-e$.
The Fermi momenta read $k_{F\sigma}=\pi n_{\sigma }$ and $k_F=\pm
[k_{F\uparrow}+ k_{F\downarrow}]/2=\pi n/2$, where
$n_{\sigma}=N_{\sigma}/N_a$ and $n=N/N_a$. The electronic density
can be written as $n=n_{\uparrow }+n_{\downarrow}$ and the spin
density is given by $m=n_{\uparrow}-n_{\downarrow}$. In general we
consider electronic densities $n$ and spin densities $m$ in the
domains $0\leq n \leq 1$ and $0\leq m \leq n$, respectively. Our
general correlation-function expressions are derived for the
metallic phase of electronic densities $0<n<1$ and spin densities
$0<m<n$. However, taking the limit $n\rightarrow 1$ in some of
these expressions leads to correct expressions for the $n=1$
Mott-Hubbard insulator phase. Also the $m=0$ spectral function
expressions can in general be obtained by taking the limit
$m\rightarrow 0$ in our general expressions. In our applications
to the study of specific spectral functions we consider such $m=0$
expressions only.

The concept of rotated electron \cite{I,IIa,IIIb} is associated
with a unitary transformation introduced in Ref. \cite{Harris}.
For such rotated electrons, double occupation is a good quantum
number for all values of the on-site Coulombian repulsion $U$. The
electrons that occur in the 1D Hubbard model (\ref{H}) are defined
by $c_{j,\,\sigma}^{\dag}$, while the rotated electron operator
${\tilde{c}}_{j,\,\sigma}^{\dag}$ is given by,
${\tilde{c}}_{j,\,\sigma}^{\dag} =
{\hat{V}}^{\dag}(U/t)\,c_{j,\,\sigma}^{\dag}\,{\hat{V}}(U/t)$,
where ${\hat{V}}(U/t)$ is the electron - rotated-electron unitary
operator defined by Eqs. (10)-(12) of Ref. \cite{IIIb}. As a
result of the Hilbert-space electron - rotated-electron unitary
rotation, all energy eigenstates of the model are described in
terms of occupancy configurations of $s_c=1/2$ and $\sigma_c=\pm
1/2$ holons, $s_s=1/2$ and $\sigma_s=\pm 1/2$ spinons, and $c$
pseudoparticles \cite{I}. Here we denoted the $\eta$-spin and spin
projections of the quantum objects by $\sigma_c$ and $\sigma_s$,
respectively, whereas their $\eta$-spin and spin values are
denoted by $s_c$ and $s_s$, respectively. In this paper we call
the holons and spinons according to their values of $\sigma_c=\pm
1/2 $ and $\sigma_s=\pm 1/2$, respectively. We denote by
$M_{\alpha,\,\pm 1/2}$ the number of $\pm 1/2$ holons ($\alpha
=c$) or $\pm 1/2$ spinons ($\alpha =s$). The value of the number
of rotated-electron doubly occupied and unoccupied sites (and
spin-down and spin-up rotated-electron singly occupied sites)
equals that of the holon numbers $M_{c,\,-1/2}$ and $M_{c,\,+1/2}$
(and spinon numbers $M_{s,\,-1/2}$ and $M_{s,\,+1/2}$),
respectively. Thus, rotated-electron double occupation $M$ is such
that $M=M_{c,\,-1/2}$. Moreover, $M_{\alpha}
=M_{\alpha,\,+1/2}+M_{\alpha,\,-1/2}$ and $N_c=M_s$, where $N_c$
is the number of $c$ pseudoparticles. The $\pm 1/2$ holons and $c$
pseudoparticles carry charge $\pm 2e$ and $-e$, respectively,
whereas the spinons have no charge degrees of freedom. The $c\nu$
pseudoparticles (and $s\nu$ pseudoparticles) associated with
Takahasi's charge (and spin) ideal string excitations of length
$\nu$ \cite{Lieb,I} are $\eta$-spin singlet $2\nu$-holon (and spin
singlet $2\nu$-spinon) composite quantum objects
\cite{I,IIa,IIIb}. The $\pm 1/2$ holons (and $\pm 1/2$ spinons)
which are not part of such $2\nu$-holon composite $c\nu$
pseudoparticles (and $2\nu$-spinon composite $s\nu$
pseudoparticles) are called $\pm 1/2$ Yang holons (and $\pm 1/2$
HL spinons) \cite{I}. In the designations {\it HL spinon} and {\it
Yang holon}, HL stands for Heilmann and Lieb and Yang refers to C.
N. Yang, respectively, who are the authors of the papers of Ref.
\cite{HL}. We denote by $N_{\alpha\nu}$ the number of composite
$\alpha\nu$ pseudoparticles belonging to branches $\alpha =c,\,s$
and $\nu=1,2,...$. We call $L_{\alpha,\,\pm 1/2}$ the number of
$\pm 1/2$ Yang holons ($\alpha =c$) or $\pm 1/2$ HL spinons
($\alpha =s$). Note that $M_{\alpha,\,\pm 1/2}=L_{\alpha,\,\pm
1/2}+\sum_{\nu=1}^{\infty}\nu\,N_{\alpha\nu}$.

Below we are mostly interested in excited states without HL
spinons and $s\nu$ pseudoparticles such that $\nu>1$. Therefore,
for simplicity often we replace the pseudoparticle branch index
$s1$ by $s$. Thus, throughout this paper the notations $s1$ and
$s$ are equivalent. For the subspaces spanned by the ground state
and excited states generated from it by a finite number of
pseudoparticle and/or Yang holon processes, the $c$, $s$, and $c1$
pseudoparticles carry band momentum $q$ such that
$\vert\,q\vert\leq q_c^0=\pi$, $\vert\,q\vert\leq q_s^0=k_F$, and
$\vert\,q\vert\leq q_{c1}^0=[\pi -2k_F]$, respectively. This
refers to electronic densities $0<n<1$ and spin density $m=0$. For
these densities the ground state is such that there are no $-1/2$
Yang holons, the $c1$ and $s$ pseudoparticle bands are empty and
filled, respectively, and the $c$ pseudoparticles occupy
$0\leq\vert\,q\vert\leq q_{Fc}^0 =2k_F$ and thus leave
$2k_F<\vert\,q\vert\leq\pi$ empty \cite{I,IIa}. Following the
notation of Ref. \cite{IIa}, we call an {\it electron ensemble
space} a Hilbert subspace spanned by all states with fixed values
for the $N_{\uparrow}$ and $N_{\downarrow}$ electron numbers.
Furthermore, we call a {\it CPHS ensemble space}, where CPHS
stands for $c$ pseudoparticle, $-1/2$ holon, and $-1/2$ spinon, a
Hilbert subspace spanned by all states with fixed values for the
numbers $N_c$, $M_{c,\,-1/2}$, and $M_{s,\,-1/2}$. A {\it CPHS
ensemble subspace} is a Hilbert subspace spanned by all states
with fixed values for the numbers $N_c$, $L_{c,\,-1/2}$, and
$L_{s,\,-1/2}$ and for the sets of numbers $\{N_{c\nu}\}$ and
$\{N_{s\nu}\}$ corresponding to $\nu=1,2,...$ branches. We note
that one does not need to provide the values of $M_{c,\,+1/2}$ and
$M_{s,\,+1/2}$ in order to specify a CPHS ensemble space, since
these numbers are not independent \cite{IIa}. Also the numbers
$L_{c,\,+1/2}$ and $L_{s,\,+1/2}$ are not independent and one does
not need to provide these values in order to specify a CPHS
ensemble subspace.

Finally, let us introduce the useful quantum number $\iota =sgn
(q) 1=\pm 1$ which refers to the number of right pseudoparticle
movers ($\iota =+1$) and left pseudoparticle movers ($\iota =-1$).
The numbers $N_{c,\,\iota}$ of $c$ pseudoparticles and
$N_{\alpha\nu,\,\iota}$ of $\alpha\nu$ pseudoparticles of $\iota $
character are good quantum numbers. We thus introduce the $c$
pseudoparticle and $\alpha\nu$ pseudoparticle current numbers,

\begin{equation}
J_{c} ={1\over 2}\sum_{\iota =\pm 1}(\iota)\,N_{c,\,\iota} \, ;
\hspace{0.5cm} J_{\alpha\nu} ={1\over 2}\sum_{\iota =\pm
1}(\iota)\,N_{\alpha\nu,\,\iota} \, . \label{Jan}
\end{equation}
The numbers $N_{c,\,\iota}$ and $N_{\alpha\nu,\,\iota}$ can be
expressed as $N_{c,\,\iota} = N_c/2 + \iota\, J_c$ and $
N_{\alpha\nu,\,\iota} = N_{\alpha\nu}/2 + \iota\, J_{\alpha\nu}$.

Each CPHS ensemble Hilbert subspace characterized by fixed values
for the sets of numbers $N_{c}$, $\{N_{\alpha\nu}\}$ such that
$\alpha =c,s$ and $\nu =1,2,3,...$, and $\{L_{\alpha,\,-1/2}\}$
such that $\alpha =c,s$ contains different subspaces with
different values for the sets of current numbers $J_{c}$ and
$\{J_{\alpha\nu}\}$ such that $\alpha =c,s$ and $\nu =1,2,...$.
According to the notation of Ref. \cite{IIIb}, we call these
subspaces {\it J-CPHS ensemble subspaces}.

\section{THE HOLON-SPINON HAMILTONIAN AND THE UPPER HUBBARD BANDS LOWER LIMIT
SPECTRUM}

The correlation function expressions are determined by transitions
from the ground state to excited states. Let us consider that the
initial ground state belongs to a canonical ensemble space
associated with densities such that $0< n< 1$ and $0< m< n$. Often
we consider such ground states. In the application of our results
to the evaluation of specific spectral function expressions we are
mostly interested in $m=0$ initial ground states. Fortunately, we
are able to derive the corresponding $m=0$ expressions from the
general correlation functions expressions obtained for $0< m< n$.
Let us introduce the Hamiltonians ${\hat{H}}_{GL}$ and
${\hat{H}}_{HS}$ such that,
\bea
& & {\hat{H}}_{GL} = \hat{H} - {\hat{H}}_{HS} \, ; \nonumber \\
& & {\hat{H}}_{HS} = 2\mu\,{\hat{M}}_{c,\,-1/2} + 2\mu_0\,H\,
\Bigl[{\hat{M}}_{s,\,-1/2}-{\hat{N}}_{s1}\Bigr] +
\sum_{\nu=2}^{\infty}\epsilon^0_{s\nu}(0)\,{\hat{N}}_{s\nu} \, .
\label{HGL-HS}
\eea Here $\hat{H}$ is the 1D Hubbard model (\ref{H}),
${\hat{H}}_{HS}$ is the {\it holon-spinon Hamiltonian}, and in
${\hat{H}}_{GL}$ the letters GL stand for gapless. Indeed, the
energy spectrum of the Hamiltonian ${\hat{H}}_{GL}$ of Eq.
(\ref{HGL-HS}) is gapless. Moreover, ${\hat{M}}_{\alpha ,\,-1/2}$
is the $-1/2$ holon ($\alpha =c$) and $-1/2$ spinon ($\alpha =s$)
number operator given in Eqs. (24) and (25) of Ref. \cite{IIIb},
${\hat{N}}_{s\nu}$ is the $s\nu$ pseudoparticle number operator,
and $\epsilon^0_{s\nu}(q)$ is the $s\nu$ pseudoparticle energy
band defined in Refs. \cite{I,IIa,IIIb}. For $m\rightarrow 0$ and
$\nu >1$ this energy band is such that $\epsilon^0_{s\nu}(q)=0$.

Since the numbers of $-1/2$ holons, $-1/2$ spinons, and $s\nu$
pseudoparticles are good quantum numbers, the three Hamiltonians
of Eq. (\ref{HGL-HS}) commute with each other. Thus, these
Hamiltonians have the same energy eigenstates, only the
corresponding energy eigenvalues being in general different. Let
us consider an energy-$E$ eigenstate of the 1D Hubbard model. Such
a state is also an energy eigenstate of the holon-spinon
Hamiltonian ${\hat{H}}_{HS}$ whose energy eigenvalue $\omega_{HS}$
reads,

\begin{equation}
\omega_{HS} = 2\mu\,M_{c,\,-1/2} + 2\mu_0\,H\,
\Bigl[M_{s,\,-1/2}-N_{s1}\Bigr]+
\sum_{\nu=2}^{\infty}\epsilon^0_{s\nu}(0)\,N_{s\nu} \, .
\label{Om0}
\end{equation}
It is also an energy eigenstate of the Hamiltonian
${\hat{H}}_{GL}$ of Eq. (\ref{HGL-HS}) of eigenvalue
$E_{GL}=[E-\omega_{HS}]$. Analysis of the finite-size corrections
of the energy $\Delta E_{GL} =[E_{GL}-E_{GS}]\approx 0$ of all
low-energy states of the Hamiltonian ${\hat{H}}_{GL}$ reveals that
the ground state of the 1D Hubbard model is also the state of
minimal energy $\Delta E_{GL} =[E_{GL}-E_{GS}]= 0$ for such a
Hamiltonian. Indeed, for all canonical ensemble spaces the latter
Hamiltonian and the 1D Hubbard model (\ref{H}) have the same
ground state. The general spectrum of interest for the problem of
the few-electron spectral functions of the 1D Hubbard model for
excitation energies $\omega=[E-E_{GS}]=\omega_{HS} +\Delta E_{GL}$
such that $\Delta E_{GL} =[E_{GL}-E_{GS}$ is small can be written
as the sum of the finite energy $\omega_{HS}$ plus the gapless
contribution $\Delta E_{GL}$ expressed in terms of the
pseudoparticle energy bands \cite{IIa,IIIb}. This gapless
contribution is the excitation energy of the same states relative
to the Hamiltonian ${\hat{H}}_{GL}$ of Eq. (\ref{HGL-HS}).

The ground-state normal-ordered expression of an operator
$\hat{O}$ is defined as $:\hat{O}:\equiv\hat{O}-\langle
GS\vert\,\hat{O}\vert\, GS\rangle$, where $\vert\, GS\rangle$ is
the ground state. Let $\vert ex\rangle$ be any excited state. The
corresponding deviation $\Delta O$ is defined as $\Delta
O\equiv\langle ex\vert\,\hat{O}\vert\, ex\rangle-\langle
GS\vert\,\hat{O}\vert\, GS\rangle$. From now on we will deal
mostly with the two Hamiltonians (\ref{H}) and ${\hat{H}}_{GL}$ of
Eq. (\ref{HGL-HS}) while acting in the reduced Hilbert subspaces
spanned by excited states with small excitation energy $\Delta
E_{GL}$ for the Hamiltonian ${\hat{H}}_{GL}$. Thus, it is useful
to introduce the ground-state normal-ordered Hamiltonians
$:{\hat{H}}_{GL}:$ and $:\hat{H}:$ such that,

\begin{equation}
:{\hat{H}}_{GL}: = :\hat{H}: - {\hat{H}}_{HS} \, ,
\label{Hno2-nhHGL}
\end{equation}
where $\hat{H}$ is the 1D Hubbard model (\ref{H}). (Note that
since the energy eigenvalue $\omega_{HS}$ (\ref{Om0}) of the
holon-spinon Hamiltonian ${\hat{H}}_{HS}$ vanishes for the ground
state, its ground-state normal-ordered expression is such that
$:{\hat{H}}_{HS}:={\hat{H}}_{HS}$.) While the Hamiltonian
$:{\hat{H}}_{GL}:$ of Eq. (\ref{Hno2-nhHGL}) describes the gapless
part of the excitations, the holon-spinon Hamiltonian
${\hat{H}}_{HS }$ of Eqs. (\ref{HGL-HS}) and (\ref{Hno2-nhHGL})
controls the finite-energy physics. Creation of a $-1/2$ holon
requires a minimal amount of excitation energy $2\mu$. Creation of
a $-1/2$ spinon (except those which are part of $s1$
pseudoparticles) requires a minimal amount of excitation energy
$2\mu_0 H$. Creation of a $s\nu$ pseudoparticle of bare-momentum
$q$ and belonging to a $\nu>1$ branch involves creation of $\nu$
$-1/2$ spinons and is associated with an energy given by
$2\nu\mu_0 H+\epsilon^0_{s\nu}(q)$. Creation of a $c\nu$
pseudoparticle of bare-momentum $q$ involves creation of $\nu$
$-1/2$ holons and requires an amount of energy $2\nu\mu
+\epsilon^0_{c\nu}(q)$. (The pseudoparticle energy bands
$\epsilon^0_{s\nu}(q)$ and $\epsilon^0_{c\nu}(q)$ are studied in
Refs. \cite{I,IIa}.)

Although our expressions refer to general values of spin density
$m$ such that $0< m< n$, in the case of applications to
few-electron spectral functions we are most interested in
zero-magnetization. Thus, for simplicity we consider excited
states without $-1/2$ HL spinons and without $s\nu$
pseudoparticles belonging to $\nu >1$ branches. We emphasize that
finite-energy states with finite occupancies for these quantum
objects play an important role only for finite values of the spin
density. One arrives to the same final $m=0$ expressions from our
general $0<m<n$  scheme regardless we consider excited states {\it
with} or {\it without} $-1/2$ HL spinons and $s\nu$
pseudoparticles belonging to $\nu >1$ branches.

For $m\rightarrow 0$ intial ground states only excited states with
finite $-1/2$ holon occupancies have a gapped energy spectrum.
These states have a minimal finite excitation energy given by,

\begin{equation}
\omega_{HS} = E_u\,M_{c,\,-1/2} = E_u\,\Bigl[\,L_{c,\,-1/2} +
\sum_{\nu=1}^{\infty}\nu\,N_{c\nu}\Bigr] \, , \label{Om0H}
\end{equation}
Here the energy $E_u \equiv 2\mu$ is defined by Eq. (107) of Ref.
\cite{IIIb}. In the limit $m\rightarrow 0$ and for $0\leq n \leq
1$ it is an increasing function of the on-site repulsion $U$ such
that $E_u = 4t\cos (\pi n/2)$ for $U/t\rightarrow 0$ and $E_u = U+
4t\cos (\pi n)$ for $U/t\rightarrow\infty$. For any value of $U/t$
it is a decreasing function of the electronic density $n$ such
that $E_u = U + 4t$ as $n\rightarrow 0$ and $E_u$ approaches the
value of the Mott-Hubbard gap $E_{MH}$ as $n\rightarrow 1$. For
values of rotated-electron double occupation $M$ such that $M>0$,
the energy $\omega_{HS}=M\,E_u$ of Eq. (\ref{Om0H}) is finite and
corresponds to the lower limit of the $M^{th}$ upper Hubbard band.

Transitions from a ground state to excited states belonging to a
given J-CPHS ensemble subspace can be labelled by the set of
deviation numbers and deviation current numbers $\{\Delta
N_{\alpha}\}$, $\{\Delta J_{\alpha}\}$, $\{\Delta
N_{c\nu}\}=\{N_{c\nu}\}$, $\{\Delta J_{c\nu}\}=\{J_{c\nu}\}$, and
$\{\Delta L_{c,\,-1/2}\}=\{L_{c,\,-1/2}\}$ where $\alpha=c,\,s$
and $\nu=1,\,2,...$. Since there are no $c\nu$ pseudoparticles and
$-1/2$ Yang holons in the ground state, below we replace the
deviations $\Delta N_{c\nu}$, $\Delta J_{c\nu}$, and $\Delta
L_{c,\,-1/2}$ by the corresponding number and current number
values $N_{c\nu}$, $J_{c\nu}$, and $L_{c,\,-1/2}$, respectively.
All states belonging to the same J-CPHS ensemble subspace have the
same number of $-1/2$ holons and thus the same energy eigenvalue
$\omega_{HS}=M\,E_u$ relative to the holon-spinon Hamiltonian
${\hat{H}}_{HS}$ of Eq. (\ref{HGL-HS}). We are mostly interested
in the {\it reduced J-CPHS ensemble subspaces}. Such subspaces are
the part of the general J-CPHS ensemble subspaces which is spanned
by states of excitation energy $\omega$ (relative to the 1D
Hubbard model) such that $\Delta E_{GL}=(\omega -M\,E_u)$ is
small. Transitions from a ground state to these states correspond
to transitions to the lower limit of the $M^{th}$ upper Hubbard
band. Importantly, the $c\nu$ pseudoparticle creation processes
which generate the $M_{c,\,-1/2}=M=1,2,...$ excited states of
vanishing energy $\Delta E_{GL}\approx 0$ and $M>0$ excitation
energy $\Delta E=\omega\approx M\,E_u$ for the 1D Hubbard model
involve $c\nu$ pseudoparticles of bare-momentum value $q\approx
\pm q^0_{c\nu} = \pm[\pi -2k_F]$ only. In addition, these states
can involve  $-1/2$ Yang holons creation and low-energy $c$ and
$s$ pseudoparticle creation and/or annihilation and particle-hole
pseudoparticle processes in the vicinity of the corresponding $c$
and $s$ pseudoparticle {\it Fermi points}. Moreover, since we are
considering final states with no $-1/2$ HL spinons and no $s\nu$
pseudoparticles belonging to $\nu>1$ branches, the number
deviations and numbers $\Delta N_c$, $\Delta N_s$, $L_{c,\,-1/2}$,
and $\{N_{c\nu}\}$ where $\nu=1,\,2,...$ obey the sum rules (51)
and (52) of Ref. \cite{IIa} with $\Delta L_{s,\,-1/2}=0$ and
$\Delta N_{s\nu} =0$ for $\nu >1$ (In that reference $N_{s\nu}$ is
denoted by $N_{s,\,\nu}$.)

When defined in a reduced J-CPHS ensemble subspace, the momentum
operator (\ref{Popel}) can be written in normal order relative to
the ground state as follows,
\bea
& & :\hat{P}:= {\hat{P}}_0 + :{\hat{P}}_{GL}: \, ; \nonumber \\
& & {\hat{P}}_0 = \sum_{\alpha =c,s} q^0_{F\alpha}\,2\,
:{\hat{J}}_{\alpha}: + \sum_{\nu
=1}^{\infty}2k_F\,2\,{\hat{J}}_{c\nu} + \pi\,{\hat{M}}_{c,\,-1/2}
\, . \label{DPop-k0op}
\eea
Here ${\hat{P}}_0 $ is the momentum associated with the
holon-spinon Hamiltonian ${\hat{H}}_{HS }$ of Eq. (\ref{HGL-HS}).
Moreover, the ground-state normal-ordered operator
$:{\hat{P}}_{GL}:$ of Eq. (\ref{DPop-k0op}) is the momentum
operator associated with the Hamiltonian $:{\hat{H}}_{GL}:$ of Eq.
(\ref{Hno2-nhHGL}) and reads,
\bea
& & :{\hat{P}}_{GL}: = {2\pi\over N_a}\,\sum_{\alpha
=c,s}\,:{\hat{N}}_{\alpha}:\Bigl[ :{\hat{J}}_{\alpha}: -
\delta_{\alpha,\,c}\,\sum_{\nu=1}^{\infty}{\hat{J}}_{c\nu}\Bigr] +
{\hat{P}}_{ph} \, ; \nonumber \\
& & {\hat{P}}_{ph} = {2\pi\over
N_a}\,\sum_{\alpha =c,s}\sum_{\iota =\pm 1}\iota\,
{\hat{N}}^{ph}_{\alpha,\,\iota}\, . \label{PGLop-Pph}
\eea
The operator ${\hat{P}}_{ph}$ is such that
${\hat{N}}^{ph}_{\alpha,\,\iota}$ counts the number
$N^{ph}_{\alpha,\,\iota}$ of momentum $\iota\,[2\pi/N_a]$
elementary particle-hole pseudoparticle processes around the
$\alpha=c$ and $\alpha=s$ {\it Fermi points}
$\iota\,q^0_{Fc}=\iota\,2k_F$ and
$\iota\,q^0_{Fs}=\iota\,k_{F\downarrow}$, respectively.

The momentum operators involved in Eq. (\ref{DPop-k0op}) commute
with each other and with the three Hamiltonians appearing in Eq.
(\ref{Hno2-nhHGL}). Thus, the energy eigenstates are also
eigenstates of the latter momentum operators. The corresponding
momentum eigenvalues are such that,

\begin{equation}
\Delta P = k_M^l + \Delta P_{GL} \, ; \hspace{0.5cm} k_M^l =
\sum_{\alpha =c,\,s} q^0_{F\alpha}\,2\,\Delta J_{\alpha} +
\,\sum_{\nu =1}^{\infty}2k_F\,2\,J_{c\nu} + \pi\,M_{c,\,-1/2} \, .
\label{k0}
\end{equation}
The index $l$ labels the different momentum values $k_M^l$
occurring for the same value of rotated-electron double occupation
$M$, as illustrated in the applications provided below. The
momentum $\Delta P_{GL}$ of Eq. (\ref{k0}) reads,

\begin{equation}
\Delta P_{GL} = {2\pi\over N_a}\,\Bigl[\,\Delta N_{c}\,[\Delta
J_{c}-\sum_{\nu=1}^{\infty} J_{c\nu}]+ \Delta N_{s}\,\Delta J_{s}
+ \sum_{\iota =\pm 1}\sum_{\alpha =c,\,s}\iota\,
N^{ph}_{\alpha,\,\iota}\Bigr] \, . \label{PGL}
\end{equation}
The current number deviations $\Delta J_{c}$, $\Delta J_{s}$, and
$J_{c\nu} =\Delta J_{c\nu}$ of Eqs. (\ref{k0}) and (\ref{PGL}) are
given by,
\bea
& & \Delta J_{c} = \Bigl({\Delta N_c + \Delta N_s +
\sum_{\nu=1}^{\infty}N_{c\nu}\over 2}\Bigr)\,{\rm mod 1} \, ;
\hspace{0.5cm} \Delta J_{s} = \Bigl({\Delta N_c\over
2}\Bigr)\,{\rm mod 1} \, ; \nonumber \\
& &  J_{c\nu} = {1\over
2}\sum_{\iota =\pm 1}\,\iota\,N_{c\nu,\,\iota} \, . \label{DJcDJs}
\eea

The momentum values $k_M^l$ defined in Eq. (\ref{k0}) play the
same role for the J-CPHS ensemble subspaces as the energies
$\omega_{HS}=M\,E_u$ for the CPHS ensemble spaces. While the
straight horizontal line $(k,\,\omega=M\,E_u)$ defines the lower
limit of the $M^{th}$ upper Hubbard band, the points $(k=
k_M^l,\,\omega= M\,E_u)$ of the same line define the edges of the
$M^{th}$ upper Hubbard band, as further discussed in Sec. IV. We
call ${\cal H}_{red}$ the reduced J-CPHS ensemble subspaces
spanned by states of momentum $k$ and excitation energy $\omega$
such that $(k-k_M^l)$ and $(\omega -M\,E_u)$ are small. These
subspaces are characterized both by specific values for the set of
deviation numbers and numbers $\{\Delta N_{\alpha}\}$,
$\{N_{c\nu}\}$, $\{\Delta J_{\alpha}\}$, $\{J_{c\nu}\}$, and
$\{L_{c,\,-1/2}\}$ where $\alpha=c,\,s$ and $\nu=1,\,2,...$ and by
small values for $(k-k_M^l)$ and $(\omega -M\,E_u)$.

It is useful for our studies to find the specific form of the
general energy spectrum $\Delta E_{GL}$ of the Hamiltonian
$:{\hat{H}}_{GL}:$ of Eq. (\ref{Hno2-nhHGL}) for the excited
states which span a reduced subspace ${\cal H}_{red}$. That energy
spectrum can be expressed in terms of a phase-shift momentum
functional associated with pseudofermions introduced in Ref.
\cite{IIIb}. According to the results of that reference, there is
a canonical transformation that maps $\alpha$ pseudoparticles and
$c\nu$ pseudoparticles onto $\alpha$ pseudofermions and $c\nu$
pseudofermions, respectively. While the $\alpha$ pseudoparticles
and $c\nu$ pseudoparticles carry bare-momentum $q$, the $\alpha$
pseudofermions carry momentum $\bar{q}=q +Q_{\alpha} (q)/N_a$ and
the $c\nu$ pseudofermions carry momentum $\bar{q}=q +Q_{c\nu}
(q)/N_a$. The functionals $Q_{\alpha} (q)$ and $Q_{c\nu} (q)$ are
defined in Eq. (73) of Ref. \cite{IIIb}. After some algebra one
finds that for the above states $\Delta E_{GL}$ has the following
leading-order $1/N_a$ finite-size energy terms,

\begin{equation}
\Delta E_{GL} = {2\pi\over N_a}\,\sum_{\alpha =c,s}\,\sum_{\iota
=\pm 1}\, v_{\alpha}\Bigl[\,\Delta^{\iota}_{\alpha} +
N^{ph}_{\alpha,\,\iota}\Bigr] + O({1\over N_a}) \, .
\label{DEGLCF}
\end{equation}
Here $v_{\alpha}$ stands for the {\it light} group velocity
$v_{\alpha}\equiv v_{\alpha}(q^0_{F\alpha})$ where
$v_{\alpha}(q)=\partial \epsilon_{\alpha} (q)/\partial q$ for
$\alpha=c$ and $\alpha=s$. Moreover, $N^{ph}_{\alpha,\,\iota}
=0,1,2,...$ is the number of elementary particle-hole
pseudoparticle processes around the $\iota=\pm 1$ {\it Fermi
points} of the $c$ and $s$ pseudoparticle bands and
$2\,\Delta^{\iota}_{\alpha}$ stands for the following functional,

\begin{equation}
2\,\Delta^{\iota}_{\alpha} = \Bigl[{N_a\over 2\pi}\,\Bigl(\Delta
q^0_{F\alpha,\,\iota}+ {Q_{\alpha} (q^0_{F\alpha,\,\iota})\over
N_a}\Bigr)\Bigr]^2 \, ; \hspace{0.5cm} \alpha =c,\,s \, ;
\hspace{0.3cm} \iota = \pm 1 \, , \label{Dia}
\end{equation}
where $Q_{\alpha} (q)/N_a$ is the above $\alpha$ pseudofermion
momentum functional. Note that since $\bar{q}=q +Q_{\alpha}
(q)/N_a$, the functional (\ref{Dia}) is proportional to the square
of the $\alpha$ pseudofermion {\it Fermi} momentum deviation
$\Delta {\bar{q}}^0_{F\alpha,\,\iota}=\Delta
q^0_{F\alpha,\,\iota}+ Q_{\alpha} (q^0_{F\alpha,\,\iota})/N_a$. In
Appendix A it is found that for the excited states that span the
reduced J-CPHS ensemble subspaces ${\cal H}_{red}$ the functional
(\ref{Dia}) can be written as,

\begin{equation}
2\,\Delta^{\iota}_{\alpha} = \Bigl[\iota\sum_{\alpha'=c,s}\,
\xi^0_{\alpha,\,\alpha'}\,{\Delta N_{\alpha'}\over 2} +
\xi^1_{\alpha,\,c}\,[\Delta J_c -\sum_{\nu=1}^{\infty}J_{c\nu}] +
\xi^1_{\alpha,\,s}\,\Delta J_s \Bigr]^2 \, . \label{cd2}
\end{equation}
Here the parameters $\xi^j_{\alpha,\,\alpha'}$ are defined by Eq.
(\ref{xiGS-xianu}) of Appendix A.

A property with a deep physical meaning is that the finite-size
energy spectrum of the $c\nu$ pseudoparticles vanishes as the
limit $q\rightarrow\pm q^0_{c\nu} = \pm[\pi -2k_F]$ is approached.
In that limit the $c\nu$ pseudoparticles become localized and
non-interacting objects, as found in Ref. \cite{IIa} and discussed
in Appendix A. In contrast, creation of $c\nu$ pseudoparticles at
other bare-momentum values leads to finite-size $c\nu$ energy
corrections and to values of excitation energy $\omega$ such that
$\Delta E_{GL}=(\omega-M\,E_u)$ is finite. In that case the
finite-energy problem cannot be mapped onto a low-energy
conformal-field theory and our method does not apply. Behind this
vanishing of the finite-size $c\nu$ energy spectrum there is a
symmetry which imposes that the $\nu$-$-1/2$ Yang holon and $c\nu$
pseudoparticle energy and momentum spectra become the same as
$q\rightarrow \pm[\pi -2k_F]$ for one of the bare-momentum values
$-[\pi -2k_F]$ or $+[\pi -2k_F]$. Creation of $\nu$ $-1/2$ Yang
holons contributes to the 1D Hubbard model energy spectrum
$\omega=\omega_{HS}+\Delta E_{GL}$ through the finite-energy term
$\omega_{HS}=M\,E_u$ of Eq. (\ref{Om0H}) by an energy amount
$\nu\,E_u$, but does not lead to any contribution to the gapless
spectrum $\Delta E_{GL}$ (\ref{DEGLCF}) and thus to the value of
the functional (\ref{cd2}). Therefore, an excitation involving $c$
and $s$ pseudoparticle processes and creation of $\nu$ $-1/2$ Yang
holons leads to an energy spectrum of the form (\ref{DEGLCF}),
whose functional $2\,\Delta^{\iota}_{\alpha}$ is given by
(\ref{cd2}) with $\Delta J_c -\sum_{\nu=1}^{\infty}J_{c\nu}$
replaced by $\Delta J_c$. Except for this difference, the $c\nu$
pseudoparticle and $\nu$-$-1/2$ holon contributions to the energy
spectrum $\omega=\omega_{HS}+\Delta E_{GL}$ and functional
(\ref{cd2}) are the same. However, the role of the $c$
pseudoparticle current number deviation shift
$-\sum_{\nu=1}^{\infty}J_{c\nu}$ is to introduce a counter term
whose presence assures that two $M>0$ finite-energy excited states
with the same $c$ and $s$ pseudoparticle occupancies and either
one $c\nu$ pseudoparticle such that $\nu =M$ or $M$ $-1/2$ Yang
holons, respectively, have the same momentum and energy spectrum.
The same holds for general $M>0$ excited states such that
$M=\sum_{\nu}^{\infty}\nu\, N_{c\nu} +L_{c,\,-1/2}$. For
simplicity, let us consider that $M=1$. In this case the current
$-\sum_{\nu=1}^{\infty}J_{c\nu}$ reads $-J_{c1}=\mp 1/2$ and zero
for creation of a $c1$ pseudoparticle and a $-1/2$ Yang holon,
respectively. According to Eq. (\ref{DJcDJs}), for states with the
same $c$ and $s$ pseudoparticle number deviations the value of the
$c$ current number deviation $\Delta J_c$ for the one-$c1$
pseudoparticle states differs from the corresponding number of the
one-$-1/2$ Yang holon states by $\pm 1/2$. Thus, for one of the
two possible values of opposite sign of the current number
$-J_{c1}=\mp 1/2$ such difference is precisely cancelled. For that
one-$c1$ pseudoparticle state the current number $\Delta J_c
-J_{c1}$ equals the current number $\Delta J_c$ of the one-$-1/2$
Yang holon state.

\section{SPECTRAL FUNCTION EXPRESSIONS IN THE VICINITY OF THE UPPER-HUBBARD
BANDS LOWER LIMITS}

In this section we derive general finite-energy expressions for
few-electron spectral functions for excitation energy and momentum
values in the vicinity of upper-Hubbard bands lower limits.
Moreover, we apply these general expressions to the study of
specific one-electron and two-electron functions.

\subsection{THE GENERAL FINITE-ENERGY SPECTRAL FUNCTION
EXPRESSIONS}

Although for $M\,E_u>0$ the energy spectrum $\omega= M\,E_u +
\Delta E_{GL}$ of the the 1D Hubbard-model in the reduced subspace
${\cal{H}}_{red}$ refers to a finite-energy problem, the
functionals $2\Delta^{\iota}_{c}$ and $2\Delta^{\iota}_{s}$
appearing in  Eq. (\ref{DEGLCF}) equal the $c$ and $s$
primary-field dimensions of a low-energy two-component
conformal-field theory \cite{Belavin,Carmelo97pp}. We note that
the momentum spectrum (\ref{PGL}) can be expressed in terms of
these functionals as $\Delta P_{GL} = {2\pi\over
N_a}\,\sum_{\alpha =c,s}\,\sum_{\iota =\pm
1}\,\iota\,[\,\Delta^{\iota}_{\alpha} + N^{ph}_{\alpha,\,\iota}]$.
Furthermore, manipulation on the integral equations that define
the phase shifts of the $\xi^1_{\alpha,\,\alpha'}$ and
$\xi^0_{\alpha,\,\alpha'}$ expressions given in Eq.
(\ref{xiGS-xianu}) of Appendix B, reveals that those are entries
of known matrices. The two corresponding matrices are the
transpose of the dressed charge matrix and the inverse of the
transpose of the dressed charge matrix, respectively, of the
low-energy 1D Hubbard model conformal field theory
\cite{Belavin,Carmelo97pp}. This property allows the evaluation of
the 1D Hubbard-model few-electron spectral function expressions
for momentum $k$ and excitation energy $\omega$ such that both
$(k-k_M^l)$ and $(\omega-M\,E_u)$ are small. For $M\,E_u>0$ such a
1D Hubbard model finite-energy problem cannot be solved by
standard low-energy conformal field theory.

Let ${\hat{\phi}}_{\vartheta}(x,\,t)$ (and
${\hat{\phi}}^{GL}_{\vartheta}(x,\,t)$) represent an one-electron
or two-electron physical field and $\vartheta =1p,\,\rho ,\,ss$
refer for example to one-electron, charge, and singlet
superconductivity, respectively. The time evolution of
${\hat{\phi}}_{\vartheta}(x,\,t)$ (and
${\hat{\phi}}^{GL}_{\vartheta}(x,\,t)$) is described by the 1D
Hubbard Hamiltonian $:\hat{H}:$ (and Hamiltonian
$:{\hat{H}}_{GL}:$ of Eq. (\ref{Hno2-nhHGL})). The space
translations are described by the momentum operator $:\hat{P}:$ of
Eq. (\ref{DPop-k0op}) (and $:{\hat{P}}_{GL}:$ of Eq.
(\ref{PGLop-Pph})). The asymptotic expression for the low-energy
correlation function of the physical field
${\hat{\phi}}^{GL}_{\vartheta}(x,\,t)$ can be obtained by use of
conformal-field theory. Such an expression, combined with the
relation between the three Hamiltonians of Eq. (\ref{Hno2-nhHGL})
and their symmetries, is used in Appendix B to show that the
leading term in the asymptotic expansion of the corresponding
finite-energy correlation function of the 1D Hubbard model is of
the following form,

\begin{equation}
\chi_{\vartheta}(x,\,t) =\langle
GS\vert\,{\hat{\phi}}_{\vartheta}(x,\,t)\,{\hat{\phi}}_{\vartheta}(0,0)\,\vert\,GS\rangle
\propto\prod_{\alpha =c,\,s}\prod_{\iota=\pm 1}
{e^{-i[k_M^l\,x-M\,E_u\,t]}\over (x-\iota
v_{\alpha}\,t)^{2\Delta^{\iota}_{\alpha}}} \, . \label{correl}
\end{equation}
Here $\Delta^{\iota}_{\alpha}$ is the functional defined in Eq.
(\ref{cd2}) and $v_{\alpha}$ stands for the {\it light} group
velocity $v_{\alpha}\equiv v_{\alpha}(q^0_{F\alpha})$. Comparison
with the low-energy correlation-function expression (\ref{cf}) of
Appendix A for ${\hat{\phi}}^{GL}_{\vartheta}(x,\,t)$ reveals that
there is in expression (\ref{correl}) an extra phase factor
$e^{-i[k_M^l\,x-M\,E_u\,t]}$. In spite of that, for values of the
momentum $k$ and excitation energy $\omega $ such that both $(k
-k_M^l)$ and $(\omega -M\,E_u)$ are small the asymptotic of the
finite-energy correlation function is of algebraic type. The
non-interacting character of the $-1/2$ Yang holons and $c1$
pseudoparticles as the limit $q\rightarrow\pm q^0_{c\nu} = \pm[\pi
-2k_F]$ is approached justifies that the creation of such objects
is a finite-energy process that leads to the phase-factor
$e^{-i[k_M^l\,x-M\,E_u\,t]}$ of expression (\ref{correl}) only.

In order to derive the expressions of the few-electron spectral
functions one needs the imaginary part of the correlation
functions in the $k$ and $\omega $ plane. Let
$\chi_{\vartheta}(k,\,\omega)$ be the Fourier transform of the
function $\chi_{\vartheta}(x,\,t)$ given in Eq. (\ref{correl}).
The $\chi_{\vartheta}(k,\,\omega)$ expression is controlled by the
same universal exponent when one approaches the point
$(k=k_M^l,\,\omega=M\,E_u)$ from all directions $(k\rightarrow
k_M^l,\,\omega\rightarrow M\,E_u)$ in the finite spectral-weight
region corresponding to the reduced subspace ${\cal{H}}_{red}$,
except for the four lines such that $(\omega-M\,E_u) \approx\pm
v_{\alpha}\,(k-k_M^l)$. (Two lines for each $\alpha =c,s$
pseudoparticle branch -- in some cases only two of these four
lines are inside the finite spectral-weight region.) These lines
are associated with the slope at the point
$(k=k_M^l,\,\omega=M\,E_u)$ of the general $\alpha$ branch lines
studied for few-electron spectral functions in Refs.
\cite{spectral,V}. In the vicinity of these lines the $\omega$
dependence of the weight distribution is also in general of
power-law type but is controlled by exponents different from that
given below. Thus, our studies are complementary to those of Refs.
\cite{spectral,V}. For simplicity, we consider that the critical
point is approached through the line defined by $k=k_M^l$ and
$\omega\rightarrow M\,E_u $. In this case we find the following
expressions for $\chi_{\vartheta}(k,\,\omega)$,

\begin{equation}
{\rm Im}\,\chi_{\vartheta} (k_M^l ,\,\omega)\propto \Bigl(\omega -
M\,E_u\Bigr)^{\zeta_{\vartheta}} \, , \label{Ichiun}
\end{equation}
and

\begin{eqnarray}
{\rm Re}\,\chi_{\vartheta} (k_M^l ,\,\omega) & \propto &
\Bigl(\omega - M\,E_u\Bigr)^{\zeta_{\vartheta}} \, ;
\hspace{0.3cm} \zeta_{\vartheta}\neq 0 \, , \nonumber \\
& \propto & -\ln\,(\omega -M\,E_u) \, ; \hspace{0.3cm}
\zeta_{\vartheta}= 0 \, , \label{ReIchiun-zet0}
\end{eqnarray}
where the exponent reads,

\begin{equation}
\zeta_{\vartheta} = -2 + \sum_{\alpha =c,\,s}\,\sum_{\iota =\pm
1}\,2\Delta_{\alpha}^{\iota} \, , \label{zeta*}
\end{equation}
and the functional $2\Delta_{\alpha}^{\iota}$ is given in Eq.
(\ref{cd2}). Expression (\ref{Ichiun}) corresponds to the leading
term of an expansion in the small energy $(\omega - M\,E_u)$. The
exponent $\zeta_{\vartheta}$ is a rapidly increasing function of
the number of excited pseudoparticles.  The non-interacting
character of  the $-1/2$ Yang holons and $q=\pm q^0_{c\nu}$ $c\nu$
pseudoparticles justifies that for $m\rightarrow 0$ the $U/t$ and
$n$ dependence of the exponent (\ref{zeta*}) occurs through the
parameter $K_{\rho}$ of Ref. \cite{Schulz} only, as confirmed by
the exponent expressions found below. (This is not in general true
for the branch-line exponents of Refs. \cite{spectral,V}, whose
expressions involve the momentum-dependent phase shifts defined in
Ref. \cite{IIIb}.) Importantly, when for one-electron (and
two-electron) correlation functions the exponent
$\zeta_{\vartheta}$ tends to $-1$ (and $-2$) as $U/t\rightarrow 0$
or $U/t\rightarrow\infty$, expression (\ref{Ichiun}) is not valid.
In this limit the correlation-function expression is such that,

\begin{equation}
\chi_{\vartheta }(k_M^l,\,\omega) \propto {1\over \omega - M\,E_u
- i\beta} \, ; \hspace{1cm} {\rm Im}\,\chi_{\vartheta} (\pm k_M^l
,\,\omega)\propto \delta(\omega - M\,E_u) \, , \label{Ichiun-1-2}
\end{equation}
where $\beta$ is real and infinitesimal and thus the spectral
function ${\rm Im} \chi_{\vartheta }(\pm k_M^l, \omega)$ is
$\delta $-function like.

For simplicity, the general expression (\ref{Ichiun}) refers to
the vertical line corresponding to to $k=k_M^l$ and small values
of $(\omega -M\,E_u)$, yet the exponent (\ref{zeta*}) controls the
weight distribution for all other lines crossing the point $(k=
k_M^l,\,\omega=M\,E_u)$ except for the $\alpha$ branch lines.
Thus, although in our applications to the one-electron addition
spectral function, dynamical structure factor, and singlet Cooper
pair addition spectral function we use the $k=k_M^l$ expression
(\ref{Ichiun}), we emphasize that similar expressions apply to the
weight distribution associated with momentum values $k$ and
excitation energy $\omega$ such that $(k -k_M^l)$ and $(\omega
-M\,E_u)$ are small.

Except in particular limits of parameter space, our present method
does not provide the values of the constants that multiply the
weight-distribution power-law expressions. This justifies the use
of the proportionally symbol $\propto$ in the general expression
(\ref{Ichiun}). Fortunately, there is a relation between the value
of the critical exponents that control the weight distributions in
the vicinity of the different upper Hubbard band lower-limit
points and the relative value of the corresponding power-law
multiplicative constants. In general, the smallest is the critical
exponent, the largest is the value of the corresponding
multiplicative constant. Such a relation is confirmed by analysis
of the results obtained for $U/t\rightarrow\infty$ by the method
of Ref. \cite{Penc97}. The finite-$U/t$ numerical results of Ref.
\cite{Senechal} and the small $U/t$ results of Ref.
\cite{opticalthey} also confirm that relation. Moreover, often the
form of the weight distributions at $U/t=0$ also provides an
useful boundary condition. Thus, our considerations for finite
values of $U/t$ concerning the relative amount of spectral weight
located in the vicinity of the different upper Hubbard band
lower-limit points result from comparison with the known limiting
results, including those given in the above references.

For $M\,E_u=0$ our correlation-function expressions reduce to the
well known conformal-field general expressions \cite{Belavin}.
However, for $M\,E_u>0$ our expressions provide new useful
information about the finite-energy spectral properties of the 1D
Hubbard model.

\subsection{INTRODUCTION TO THE FEW-ELECTRON FINITE-ENERGY PROBLEM}

All our applications refer to positive excitation energy $\omega$.
In this case the general correlation function
$\chi_{\vartheta}(k,\,\omega)$ can be written in terms of a
Lehmann representation as follows,

\begin{equation}
\chi_{\vartheta }(k,\,\omega) = \sum_{j} {\vert\langle j\vert
{\hat{\cal{O}}}_{\vartheta } (k) \vert GS\rangle\vert^2 \over
\omega -\omega_{j,\,0} - i\beta} \, , \label{corrpos}
\end{equation}
where $\beta\rightarrow 0$ and the operator
${\hat{\cal{O}}}_{\vartheta } (k)$ is the Fourier transform of the
general one-electron or two-electron physical field
${\hat{\phi}}_{\vartheta}(x,\,0)$ of Eq. (\ref{t}) of Appendix B.
The $j$ summations run now over all available final excited energy
eigenstates and $\omega_{j,\,0}= [E_{i}-E_{GS}]$ are the
Hubbard-model excitation energies relative to the initial ground
state.

In our applications of the general spectral-function expressions
we consider the following operators ${\hat{\cal{O}}}_{\vartheta }
(k)$: The spin-up one-electron addition operator, the charge
operator, and the singlet-superconductivity Cooper-pair addition
operator,
\bea
& & {\hat{\cal{O}}}_{1p} (k) = c^{\dagger}_{k,\,\uparrow} \, ;
\hspace{0.5cm} {\hat{\cal{O}}}_{\rho} (k)  =
\sum_{k'}\sum_{\sigma=\downarrow ,\uparrow}\,
c^{\dagger}_{k',\,\sigma}\, c_{k+k',\,\sigma} \, ; \nonumber \\
& & {\hat{\cal{O}}}_{ss} (k) = \sum_{k'}
c^{\dagger}_{k',\,\downarrow}\,c^{\dagger}_{k-k',\,\uparrow} \, .
\label{Ok}
\eea
As above, here we use the notations $\vartheta =1p$ for
one-electron, $\vartheta =\rho$ for charge, and $\vartheta =ss$
for $s$-wave singlet superconductivity. We are particularly
interested in the spectral functions associated with the imaginary
part of the correlation function (\ref{corrpos}),

\begin{equation}
{\rm Im}\, \chi_{\vartheta }(k,\,\omega) =
\pi\,\sum_{j}\vert\langle j \vert\, {\hat{\cal{O}}}_{\vartheta }
(k) \vert\, GS\rangle\vert^2\,\delta (\omega -\omega_{j,\,0}) \,
 ; \hspace{0.5cm} \vartheta = 1p,\,\rho ,\,ss \, . \label{Imcorrpos}
\end{equation}

We consider the case of ground states of spin density
$m\rightarrow 0$. In this case the energy $\omega_{HS} = M\,Eu$
given in Eq. (\ref{Om0H}) defines the lower limit of the $M^{th}$
upper Hubbard band. Often the momentum values $k_M^l$ given in Eq.
(\ref{k0}) and associated with the $M^{th}$ upper Hubbard band are
such that $k_M^l=[\pi -l\,k_0]$ where $k_0$ is a spectral-function
dependent momentum. In our applications we consider Hubbard bands
generated by dominant processes only. According to the results of
Ref. \cite{V}, for the spin-up one-electron addition operator
${\hat{\cal{O}}}_{1p} (k) = c^{\dagger}_{k,\uparrow}$ and charge
operator ${\hat{\cal{O}}}_{\rho} (k) =
\sum_{k'}\sum_{\sigma=\downarrow ,\uparrow}\,
c^{\dagger}_{k',\,\sigma}\, c_{k+k',\,\sigma}$ (and
singlet-superconductivity operator ${\hat{\cal{O}}}_{ss} (k)=
\sum_{k'}
c^{\dagger}_{k',\downarrow}\,c^{\dagger}_{k-k',\uparrow}$) more
than 99\% of the spectral weight corresponds to excited states
such that $M_{c,\,-1/2}=M=0,\,1$ (and $M_{c,\,-1/2}=M=0,\,1,\,2$).
States with $M_{c,\,-1/2}=M>1$ (and $M_{c,\,-1/2}=M>2$) lead to
nearly no spectral weight and are omitted below. For electron
addition and the dynamical structure factor we study the weight
distribution for domains of the ($k,\,\omega$)-plane corresponding
to small values of $(k-k_1^l)$ and $(\omega -E_u)$ just above the
lower limit of the first upper Hubbard band. For the regions
corresponding to other momentum values and small energy $(\omega
-E_u)$ these spectral functions vanish because there is no
spectral weight. In the case of the singlet Cooper pair spectral
function we study the weight distribution both for domains of the
($k,\,\omega$)-plane corresponding to small values of $(k-k_1^l)$
and $(\omega -E_u)$ just above the lower limit of the first upper
Hubbard band and $(k-k_2^l)$ and $(\omega -2E_u)$ just above the
lower limit of the second upper Hubbard band.

\begin{figure}
\epsfxsize=10.0cm
\epsfysize=10.0cm
\centerline{\epsffile{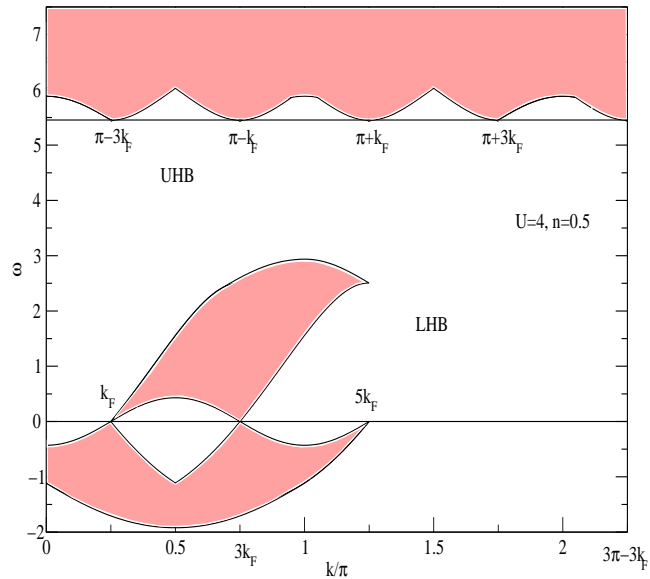}}
\begin{center}
\caption{ Regions of the ($k,\,\omega$)-plane where accordingly to
the restrictions associated with the dominant holon and spinon
microscopic processes studied in reference \cite{V} there is a
significant amount of one-electron spectral weight for on-site
repulsion $U=4t$, electronic density $n=1/2$, and spin density
$m=0$. Note that an extended momentum scheme is used. The
one-electron addition upper and lower Hubbard band spectral
functions correspond to values of excitation energy such that
$\omega>E_u$ and $0<\omega< E_u$, respectively.}
\end{center}
\label{fig1}
\end{figure}

>From analysis of the energy spectrum of the excited states
generated by dominant processes, one finds that the $M^{th}$
Hubbard band finite-spectral-weight region of the
$(k,\,\omega)$-plane whose lower limit is located at the line
$\omega =M\,E_u$ is for all values of $k$ separated from the next
lower-limit line $\omega=[M+1]\,E_u$ of the $[M+1]^{th}$ Hubbard
band by a region with nearly no spectral weight. The latter region
of the $(k,\,\omega)$-plane is out of the range of the excitations
generated by dominant processes. This holds for all few-electron
spectral functions and is illustrated in Fig. 1 for the specific
case of the one-electron spectral function. For on-site repulsion
$U=4t$, electronic density $n=1/2$, and spin density $m=0$ the
regions of the ($k,\,\omega$)-plane whose spectral weight is
generated by dominant processes are shown in the figure by shaded
areas. The dominant processes also include pseudoparticle
particle-hole processes which lead to spectral weight both inside
and outside but in the close vicinity of the shaded domains of
Fig. 1. In that figure an extended momentum scheme centered at
momentum $k=0$ (and $k=\pi$) is used for electron removal and the
$M=0$ lower Hubbard band (and $M=1$ first upper Hubbard band). The
one-electron addition upper and lower Hubbard band spectral
functions correspond to values of excitation energy such that
$\omega>E_u$ and $\omega >0$, respectively. Note that for all
values of $k$ there is indeed a region with nearly no spectral
weight located between the finite-weight regions corresponding to
the $M=0$ and $M=1$ Hubbard bands. In the figure the one-electron
removal spectral function corresponds to $\omega <0$. The figure
shows the locations ($k=\pi -l\,k_F,\,\omega= E_u$) for the four
$l=\pm 1,\pm 3$ upper Hubbard band lower-limit points whose weight
distribution is studied below. There are also two upper Hubbard
band lower-limit points for $l=\pm 5$ which are not shown in the
figure.

As in the case of the one-electron spectral weight represented in
the figure, it occurs for all few-electron spectral functions that
in the straight and horizontal lower limit upper Hubbard band
lines ($k,\,\omega=M\,E_u$) of the $(k,\,\omega$)-plane there is
finite spectral weight for a set of discrete momentum values only,
as mentioned above. These discrete momentum values coincide with
the momenta $k_M^l$ given in Eq. (\ref{k0}). The four momenta
$k=\pi-l\,k_F$ where $l=\pm 1,\,\pm 3$ that are shown in Fig. 1
are examples of such momentum values. Each of these momentum
values correspond to a different reduced J-CPHS ensemble subspace
of the CPHS ensemble space spanned by states of rotated electron
double occupation $M$ and with an occupancy of $M=M_{c,\,-1/2}$
$-1/2$ holons. Our scheme provides the weight distribution for
small energies $(\omega -M\,E_u)$ corresponding to regions of the
($k,\,\omega$)-plane just above such straight and horizontal
lower-limit upper Hubbard band lines. For the straight and
horizontal lines parallel to the lower-limit lines and
corresponding to values of $\omega$ such that $(\omega -M\,E_u)$
is small, there is finite spectral weight for small momentum
domains centered around the set of momentum values $k=k_M^l$.

The correlation-function expressions obtained from standard
conformal-field theory \cite{Belavin,Carmelo97pp} correspond to a
specific limit of our scheme such that $M=0$. The corresponding
low-energy edges result from ground-state transitions to
$M_{c,\,-1/2}=M=0$ excited states. In figure 1 this corresponds to
the finite-weight regions in the vicinity of the three points
($k=k_F,\,\omega=0$), ($k=3k_F,\,\omega=0$), and
($k=5k_F,\,\omega=0$). In the following applications of our
finite-energy method we do not consider weight distributions
resulting from such low-energy transitions, which can be evaluated
by means of two-component conformal-field theory
\cite{{Carmelo97pp}}. Our studies are restricted to the weight
distribution of each of the above spectral functions in the
vicinity of finite-energy points associated creation of $-1/2$
Yang holons and/or $c\nu$ pseudoparticles of bare momentum
$q\approx \pm q^0_{c\nu} = \pm[\pi -2k_F]$. The weight
distributions resulting from creation of $c\nu$ pseudoparticles at
other bare-momentum values appear at finite energy $\omega$ such
that $(\omega -\omega_{HS})$ is finite. These weight distributions
cannot be evaluated by our method and are studied elsewhere
\cite{spectral0,spectral,V}. Below we find that several critical
exponents which control the weight distribution in the vicinity of
the first and second upper Hubbard band lower-limit points equal
corresponding exponents that control the low-energy weight
distributions studied in Ref. \cite{Carmelo97pp}. However, we note
that in the figures that reference these exponents were plotted
for finite values of the magnetic field whereas our study refers
to zero field and magnetization.

Transitions from the ground state to excited states belonging to
the same J-CPHS ensemble subspace are of the {\it same type},
whereas transitions of {\it different type} refer to excited
states belonging to different J-CPHS ensemble subspaces. In some
cases we extend this concept to pairs of J-CPHS ensemble subspaces
with the same absolute momentum value for $k^l_M$.

\subsection{ONE-ELECTRON ADDITION UPPER-HUBBARD BAND WEIGHT DISTRIBUTION}

We start by considering addition of a spin-up electron. According
to Eq. (\ref{DJcDJs}) and Eqs. (51) and (52) of Ref. \cite{IIa}
with $\Delta L_{s,\,-1/2}=0$ and $\Delta N_{s\nu} =0$ for $\nu
>1$, in the case of the $M_{c,\,-1/2}=M=1$ type of transitions one has
that $\Delta N_c =\Delta N_s =-1$ and $\Delta J_s=\pm 1/2$. There
are two one-$-1/2$ Yang holon types of transitions and six one-$q=
\pm q^0_{c1} = \pm[\pi -2k_F]$ $c1$ pseudoparticle types of
transitions. For $m\rightarrow 0$ the number of momentum values
$k=k_1^l$ associated with dominant processes such that
$\omega=E_u$ is in the present case six. These momentum values
read $\pi- l\,k_F$ where $l=\pm 1,\,\pm 3,\,\pm 5$. Indeed, two
one-$-1/2$ Yang holon type of transitions and two of the six
one-$q= \pm q^0_{c1} = \pm[\pi -2k_F]$ $c1$ pseudoparticle type of
transitions have the same momentum and energy spectrum. Thus, they
contribute to the same upper Hubbard band lower-limit weight
distribution located around the points $(k=\pi \mp k_F,\,\omega
=E_u)$. These are the two one-$-1/2$ Yang holon types of
transitions such that $L_{c,\,-1/2}=1$, $\Delta J_c =0$, and
$\Delta J_s=\pm 1/2$ and the two one-$q= \pm q^0_{c\nu} = \pm[\pi
-2k_F]$ $c1$ pseudoparticle types of transitions such that
$N_{c1}=1$, $\Delta J_c =\pm 1/2$, $J_{c1}=\pm 1/2$, and $\Delta
J_s=\pm 1/2$. Two other upper Hubbard band lower-limit points are
($k=\pi \mp 3k_F$,\,$\omega =E_u$). The corresponding spectral
weight is generated by types of transitions such that $N_{c1}=1$,
$\Delta J_c =\pm 1/2$, $J_{c1}=\mp 1/2$, and $\Delta J_s=\mp 1/2$.
Below we also study the weight distribution in the vicinity of
these points. In their vicinity there is a smaller amount of
spectral weight than in the vicinity of the above two former
points. Finally, the weight located in the vicinity of the
remaining two points is even smaller. It is generated by
transitions such that $N_{c1}=1$, $\Delta J_c =\pm 1/2$,
$J_{c1}=\mp 1/2$, and $\Delta J_s=\pm 1/2$. These two points are
located at ($k=\pi \mp 5k_F$,\,$\omega =E_u$). In the following we
study the weight distribution around the above six upper Hubbard
band lower-limit points located at ($k=\pi -l\,k_F,\,\omega= E_u$)
where $l=\pm 1,\,\pm 3,\,\pm 5$.

Let us consider the limit $m\rightarrow 0$ where the spin-up and
spin-down one-electron spectral functions have the same form. Use
of the general expression (\ref{Ichiun}) leads for excitation
energy $\omega$ just above $E_u$ and momentum values $k=\pi
-l\,k_F$ where $l=\pm 1,\,\pm 3,\,\pm 5$ to,

\begin{equation}
{\rm Im}\,\chi_{1p} (\pi -l\,k_F,\,\omega)\propto (\omega -
E_u)^{\zeta_{1p}^{\vert l\vert}} \, ; \hspace{0.5cm} l=\pm 1,\,\pm
3,\,\pm 5 \, . \label{Ichiun1p}
\end{equation}
In this case the general exponent (\ref{zeta*}) is given by,

\begin{equation}
\zeta_{1p}^{\vert l\vert} =  -{3\over 2} +  {1\over
2}\,\Bigl[\,[1/\sqrt{2K_{\rho}}]^2+[l\,\sqrt{2K_{\rho}}/2]^2\Bigr]
\, ; \hspace{0.5cm} l=\pm 1,\,\pm 3,\,\pm 5 \, . \label{zetasGpm0}
\end{equation}

To reach this result we first evaluated the expression for general
values of $m$ in terms of the parameters (\ref{xiGS-xianu}) of
Appendix A. Use of the limiting values for these parameters given
in the same Appendix then leads to expression (\ref{zetasGpm0}).
The same procedure is followed in the calculation of the
$m\rightarrow 0$ weight distribution of the dynamical structure
factor and singlet Cooper pair spectral function given below. Note
that in the present case all the transitions leading to the weight
distributions (\ref{Ichiun1p}) involve creation of both a $c$
pseudoparticle hole and a $s$ pseudoparticle hole. Note also that
the dependence on $n$ and $U/t$ of the exponent (\ref{zetasGpm0})
occurs through the parameter $K_{\rho}$ only, as mentioned in the
previous section. For $m=0$ this is a general property of the
critical exponents that control the weight distribution in the
vicinity of the upper-Hubbard bands lower limit. In contrast, the
expressions of the exponents that control the weight distribution
in the vicinity of the branch lines studied in Ref.
\cite{spectral} involve the momentum-dependent phase shifts
defined in Ref. \cite{IIIb}.

The exponent (\ref{zetasGpm0}) is a function of $U/t$ which for
electronic densities $n$ such that $0<n<1$ changes from
$\zeta_{1p}^{\vert l\vert}\rightarrow (l^2-5)/4=-1,\,1,\,5$ for
$l=\pm 1,\,\pm 3,\,\pm 5$, respectively, as $U/t\rightarrow 0$ to
$\zeta_{1p}^{\vert l\vert}\rightarrow (l^2-8)/8=-7/8,\,1/8,\,17/8$
for $l=\pm 1,\,\pm 3,\,\pm 5$, respectively, as
$U/t\rightarrow\infty$. Note that as the critical exponent
(\ref{zetasGpm0}) approaches $-1$ for $l=\pm 1$ as $U/t\rightarrow
0$ the spectral function behaves as given in Eq.
(\ref{Ichiun-1-2}). The exponents (\ref{zetasGpm0}) are plotted in
Figs. 2 (a) and (b) for $k=\pi\mp k_F$ and $k=\pi\mp 3k_F$,
respectively, as a function of the electronic density $n$ and for
different values of $U/t$. The $l=\pm 5$ exponent is not plotted
in the figure. This exponent is larger than two and corresponds to
regions of very little spectral weight. The exponents plotted in
Fig. 2 are monotonous functions of $n$. The exponent
$\zeta_{1p}^1$ is always negative and such that $-1\leq
\zeta_{1p}^1\leq -7/8$ and is associated with a
weight-distribution singularity. It increases for increasing
values of $U/t$. For finite values of $U/t$ it is a function of
the electronic density $n$ with a minimum for an intermediate
value of $n$. The exponent $\zeta_{1p}^3$ is always positive and
such that $1/8\leq \zeta_{1p}^3\leq 1$ and is associated with a
weight-distribution edge. The exponent $\zeta_{1p}^3$ decreases
for increasing values of $U/t$. For finite values of $U/t$ it is a
function of the electronic density $n$ with a maximum for an
intermediate value of $n$.

\begin{figure}
\epsfxsize=10.0cm \epsfysize=10.0cm
\centerline{\epsffile{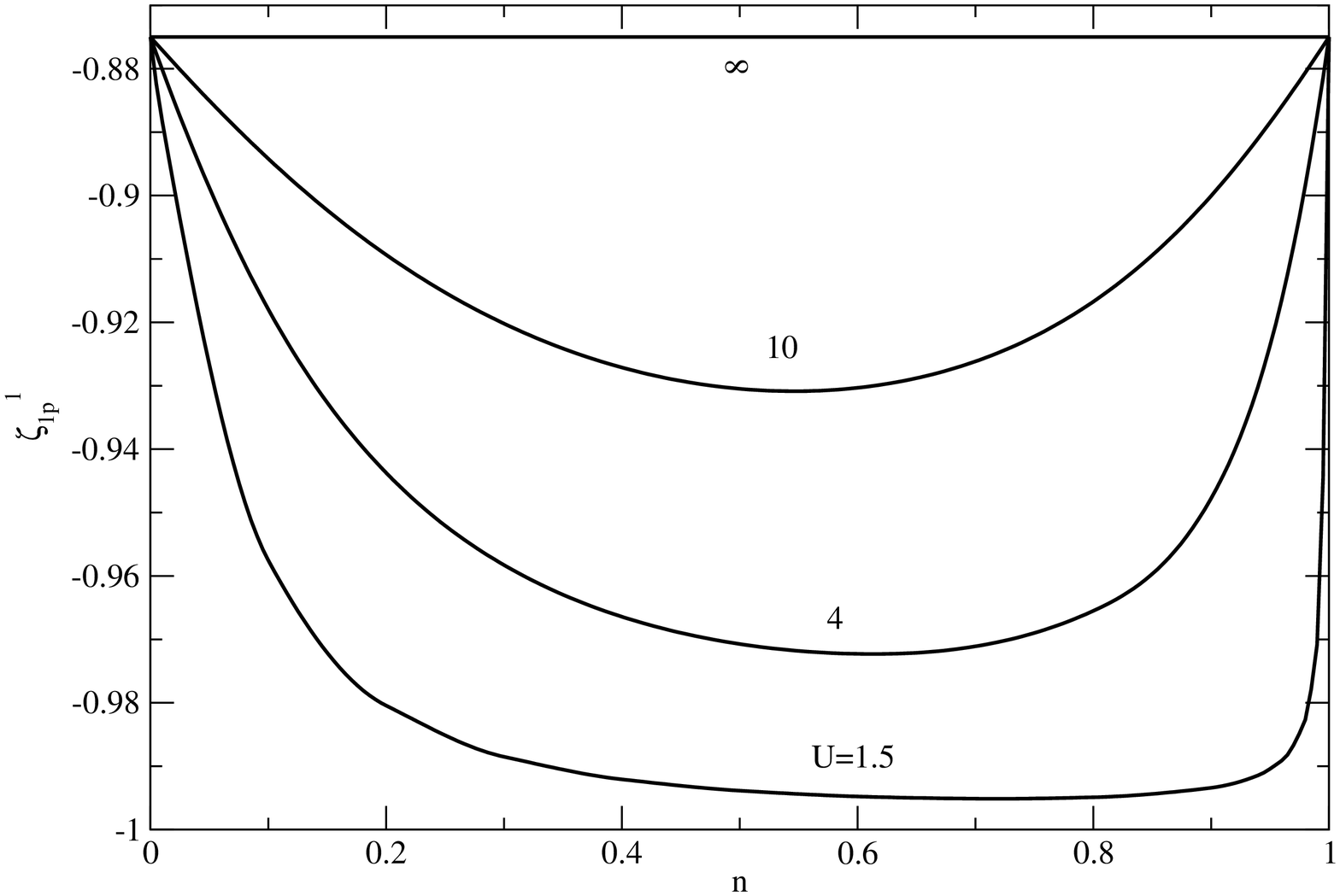}} \epsfxsize=10.0cm
\epsfysize=10.0cm \centerline{\epsffile{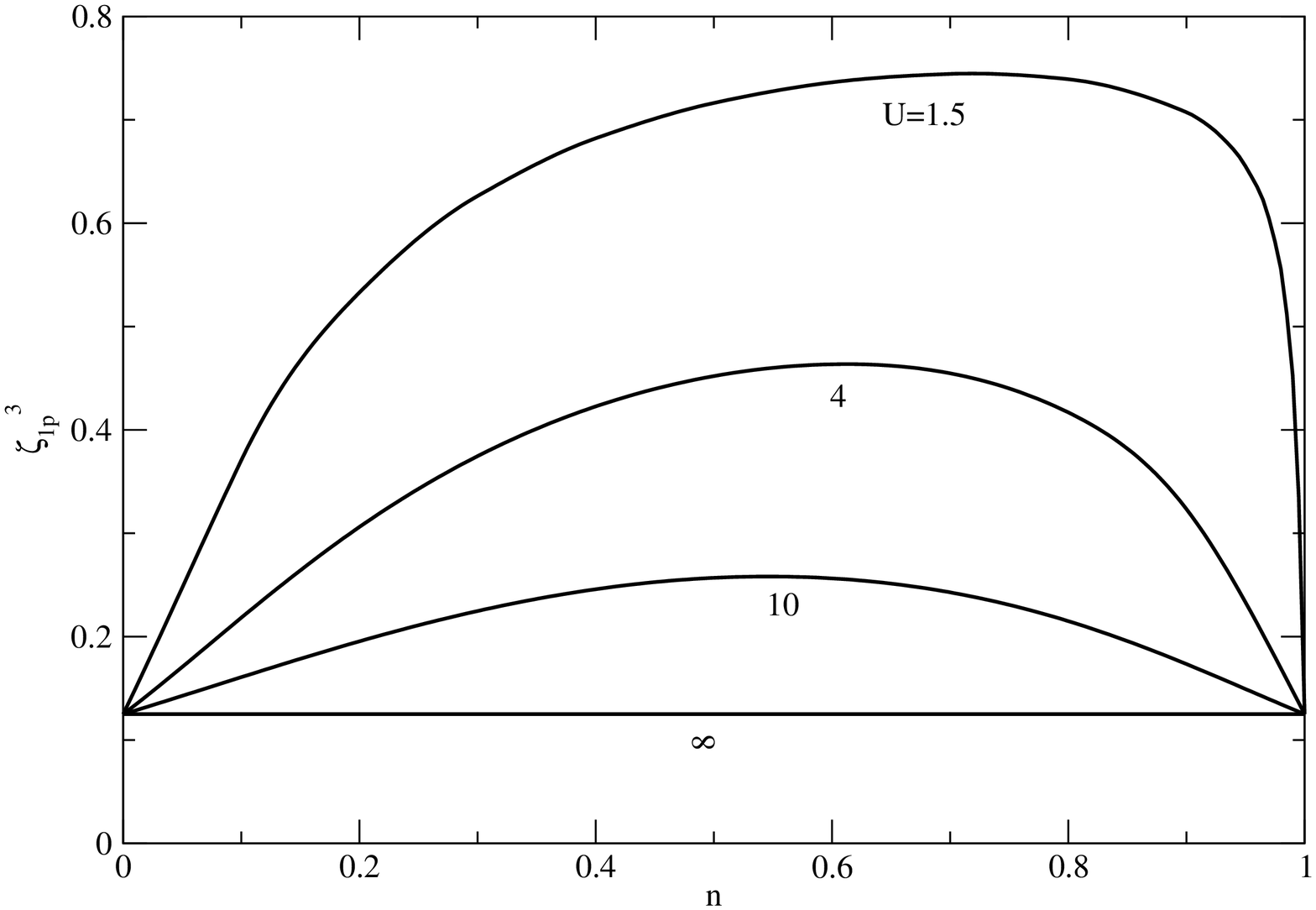}}
\begin{center}
\caption{ The upper Hubbard band one-electron exponents (a)
$\zeta_{1p}^1$ and (b) $\zeta_{1p}^3$ given in Eq.
(\ref{zetasGpm0}) as a function of the electronic density $n$ for
$U=1.5t$, $U=4t$, $U=10t$, and $U\rightarrow\infty$ (horizontal
lines).}
\end{center}
\label{fig2}
\end{figure}

For all values of $U/t$ there is more spectral weight around the
lower-limit upper Hubbard band points located at ($k= \pi \mp
k_F,\,\omega= E_u$) than around the points located at ($k=\pi \mp
3k_F,\,\omega= E_u$). Moreover, the amount of weight in the
vicinity of the points located at ($k=\pi \mp 3k_F,\,\omega= E_u$)
is larger than that of the weight located in the vicinity of the
points ($k=\pi \mp 5k_F,\,\omega= E_u$). For $U/t\rightarrow 0$
the spectral weight located in the vicinity of the points ($k=\pi
\mp 3k_F,\,\omega= E_u$) and ($k=\pi \mp 5k_F,\,\omega= E_u$)
disappears. As mentioned above, the value of the critical exponent
plotted in Fig. 2 (a) is $-1$ for $U/t\rightarrow 0$. Thus, in
this limit the spectral function behaves as given in expression
Eq. (\ref{Ichiun-1-2}). Importantly, in that limit the energy
$E_u$ of the upper Hubbard band lower limit equals the
non-interacting electronic spectrum at ($k=\pi \mp k_F,\,\omega=
E_u$). It follows that our general $U/t$ weight distribution leads
to the correct non-interacting spectrum in the $U/t\rightarrow 0$
limit. In this paper we study the weight distribution in the
vicinity of points located in the lower limit of the Hubbard
bands. The branch-line studies of Ref. \cite{spectral} reveal that
the lower-Hubbard band and upper-Hubbard band spectral weights
become connected at $k=\pi -k_F$ as $U/t\rightarrow 0$. For finite
values of $U/t$, the finite-weight regions of these two bands are
separated by a region out of the range of dominant processes. This
is confirmed by analysis of Fig. 1.

Our above finite-energy expressions were derived for the metallic
phase corresponding to electronic densities $0<n<1$. In the limit
$n\rightarrow 1$ and finite values of $U/t$ expressions
(\ref{zetasGpm0}) lead to the correct $n=1$ results. In that limit
and for $m\rightarrow 0$ the upper Hubbard band lower-limit points
($k=\pi\mp k_F,\,\omega = E_u$) coincide with the lower-limit
points ($k=\pi\mp 3k_F,\,\omega = E_u$), and ($k_=\pi\mp
5k_F,\,\omega = E_u$). Thus, in this case the smallest of the
three exponents (\ref{zetasGpm0}), such that $l=1$, controls the
leading-order weight-distribution term (\ref{Ichiun1p}).

In the limit $n\rightarrow 1$ the equality of the exponent that
controls the electron removal lower-Hubbard band and the $l=\pm 1$
finite-energy electron addition exponent (\ref{zetasGpm0}) is
required by particle-hole symmetry. Our results reveal that in the
metallic phase the $l=\pm 1$, $l=\pm 3$, and $l=\pm 5$
finite-energy exponents (\ref{zetasGpm0}) also equal the exponents
that control the weight distribution in the vicinity of the three
zero-energy points located at ($k=\pm k_F,\,\omega = 0$), ($k=\pm
3k_F,\,\omega = 0$), and ($k=\pm 5k_F,\,\omega = 0$),
respectively. (The points ($k=k_F,\,\omega= 0$), ($k=3k_F,\,\omega
= 0$), and ($k=5k_F,\,\omega = 0$) are shown in Fig. 1.) For
$l=\pm 1$ the exponent expression (\ref{zetasGpm0}) was already
found in Ref. \cite{Carmelo97pp}. It corresponds to the
$H\rightarrow 0$ expressions of the one spin-up and spin-down
electron exponents given in Table I of that reference, where $H$
is the magnetic field.

There are not many previous results about the excitation energy
and momentum dependence of the weight distribution in the vicinity
of the UHB lower-limit points considered above. Conformal-field
theory does not apply to finite energy. The method used in Ref.
\cite{Penc97} refers to $U/t\rightarrow\infty$ where the UHB lower
limit corresponds to $\omega =E_u=\infty$. Thus, the UHB was not
considered in the finite-energy studies of that reference. The
point ($k=\pi -\,k_F,\,\omega= E_u$) corresponds to
($k=\pi/2,\,(\omega -\mu)/t\approx 0.5$) and ($k=7\pi/12\approx
0.58\,\pi,\,(\omega -\mu)/t\approx 1$) in Figs. 3 and 4 of Ref.
\cite{Senechal}, respectively. The spectral weight of Fig. 3 is
for half filling and $U/t=4$. (The figure zero-energy is the
middle of the Mott-Hubbard gap.) The weight of Fig. 4 is for
electronic density $n=5/6$ and $U/t=4$. These weight distributions
were obtained by numerical calculations based on a combination of
exact diagonalizations of finite clusters with strong-coupling
perturbation theory and are consistent with our results.
Unfortunately, a quantitative comparison is not possible because
the method used in Ref. \cite{Senechal} does not provide accurate
information about the weight-distribution dependence on the
excitation energy and momentum.

\subsection{THE DYNAMICAL STRUCTURE FACTOR UPPER-HUBBARD BAND WEIGHT DISTRIBUTION}

According to Eq. (\ref{DJcDJs}) and Eqs. (51) and (52) of Ref.
\cite{IIa} with $\Delta L_{s,\,-1/2}=0$ and $\Delta N_{s\nu} =0$
for $\nu >1$, for $M_{c,\,-1/2}=M=1$ types of transitions one has
in this case that $\Delta N_c =-2$, $\Delta N_s =-1$, and $\Delta
J_s=0$. These transitions generate spectral weight in the vicinity
of two upper Hubbard band lower-limit points. There are two
one-$-1/2$ Yang holon types of transitions and two one-$q= \pm
q^0_{c1} = \pm[\pi -2k_F]$ $c1$ pseudoparticle types of
transitions which contribute to these weight distributions. The
former two types of transitions correspond to $L_{c,\,-1/2}=1$ and
$\Delta J_c =\pm 1/2$. The latter two types correspond to
$N_{c1}=1$, $\Delta J_c =0$, and $J_{c1}=\mp 1/2$. The weight
distribution in the vicinity of the two upper Hubbard band
lower-limit points located at ($k=\pi \mp 2k_F,\, \omega =E_u$)
results from transitions belonging to these two types. Use of the
general expression (\ref{Ichiun}) leads in the limit $m\rightarrow
0$ to the following expression for the dynamical structure factor
for excitation energy $\omega$ such that $(\omega -E_u)$ is small
and momentum values $k=\pi \mp 2k_F$,

\begin{equation}
{\rm Im}\,\chi_{\rho} (\pi \mp 2k_F,\,\omega)\propto (\omega -
Eu)^{\zeta_{\rho}} \, . \label{Ichiunrho}
\end{equation}
>From use of Eq. (\ref{zeta*}), we find the following expression
for the critical exponent $\zeta_{\rho}$ in the limit
$m\rightarrow 0$,

\begin{equation}
\zeta_{\rho} = -2\Bigl[1 - [\sqrt{2K_{\rho}}/2]^2 -
[1/\sqrt{2K_{\rho}}]^2\Bigr] \, . \label{zetadsf}
\end{equation}

The exponent (\ref{zetadsf}) is plotted in Fig. 3 as a function of
the electronic density $n$ for different values of $U/t$. This
exponent is a function of $U/t$ which for electronic densities
densities $n$ such that $0<n<1$ and spin density $m=0$ changes
from $\zeta_{\rho}\rightarrow 0$ as $U/t\rightarrow 0$ to
$\zeta_{\rho}\rightarrow 1/2$ as $U/t\rightarrow\infty$. For the
$n=1$ Mott-Hubbard insulator phase the exponent reads $1/2$ for
all finite values of $U/t$. The exponent $\zeta_{\rho}$ is always
positive and such that $0\leq\zeta_{\rho}\leq 1/2$ and is
associated with a weight-distribution edge. It increases for
increasing values of $U/t$. For finite values of $U/t$ it is a
function of the electronic density $n$ with a minimum for an
intermediate value of $n$.

\begin{figure}
\epsfxsize=10.0cm \epsfysize=10.0cm
\centerline{\epsffile{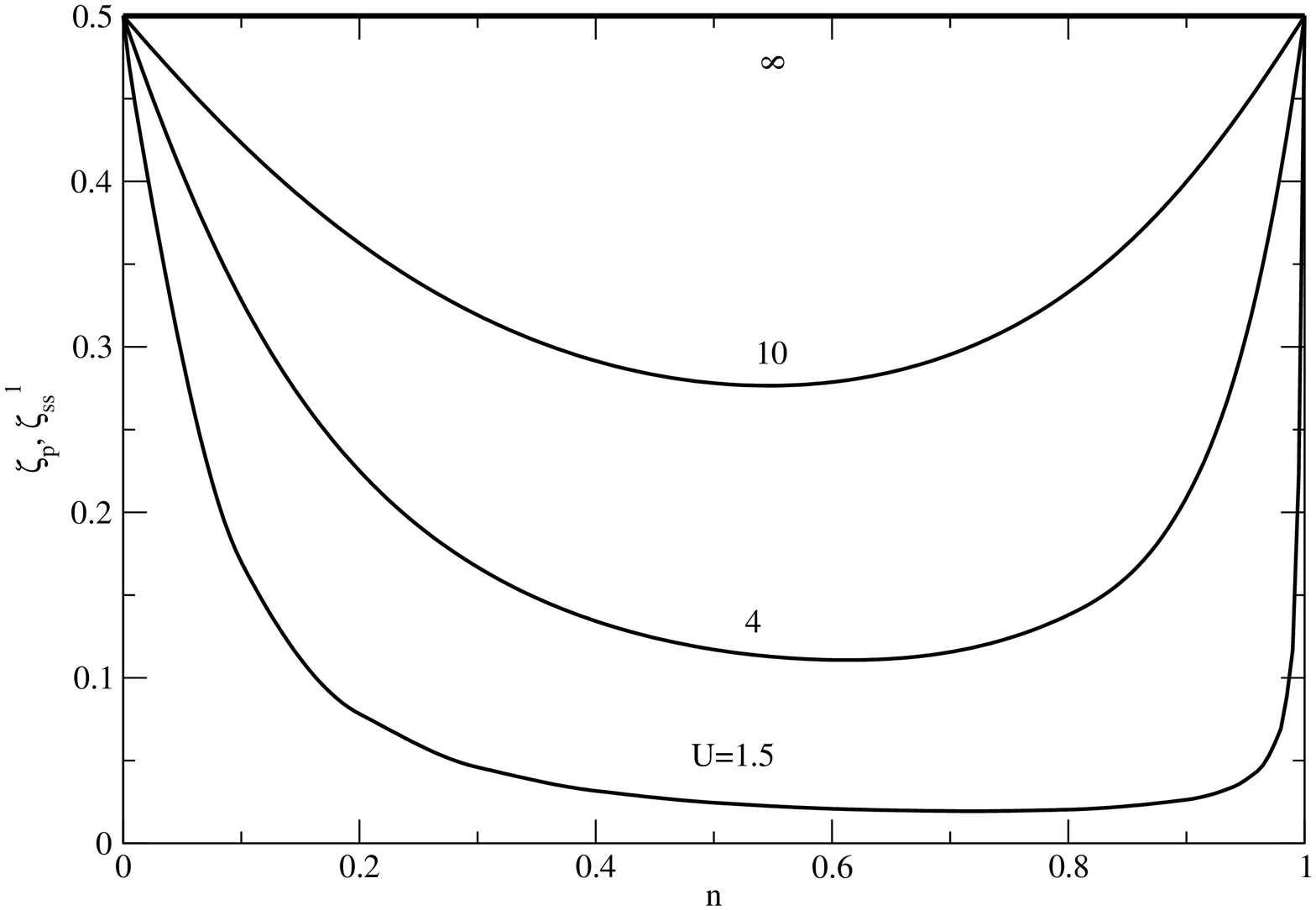}}
\begin{center}
\caption{The exponents $\zeta_{\rho}=\zeta_{ss}^1$ (\ref{zetadsf})
that control the weight distribution of the upper Hubbard band
dynamical structure factor and second upper Hubbard band singlet
Cooper pair spectral function, respectively, as a function of the
electronic density $n$ for $U=1.5t$, $U=4t$, $U=10t$, and
$U\rightarrow\infty$.}
\end{center}
\label{fig3}
\end{figure}

In contrast to the above one-electron problem, the spectral weight
associated with the dynamical structure factor upper Hubbard band
vanishes in the limit $U/t\rightarrow\infty$. For the $n=1$
Mott-Hubbard transition, the whole dynamical structure factor
vanishes as $U/t\rightarrow\infty$. This behavior is related to
the vanishing of the kinetic energy as $U/t\rightarrow\infty$
\cite{Carmelo86}. The dynamical structure factor upper Hubbard
band exponent (\ref{zetadsf}) had not been studied until now. It
controls an onset of spectral weight whose $\omega$ derivative is
infinite at the lower limit of the upper Hubbard band $\omega=E_u$
and raises upwards according to the power law (\ref{Ichiunrho}).

In general the momentum values $k=\pi \mp 2k_F$ are finite and
thus in the metallic phase the weight distribution
(\ref{Ichiunrho}) is not related to the zero-momentum frequency
dependent optical conductivity. However, in the limit
$n\rightarrow 1$ these momentum values vanish. Thus, for the
particular case of the Mott-Hubbard insulator one can use the
following relation between ${\rm Im}\,\chi_{\rho} (k,\,\omega)$
and the regular part of the frequency dependent optical
conductivity $\sigma^{reg} (\omega)$,

\begin{equation}
{\rm Re}\,\sigma^{reg} (\omega)\propto \lim_{k\rightarrow
0}{\omega\, {\rm Im}\,\chi_{\rho} (k,\,\omega) \over k^2} \, ,
\label{condDSSF}
\end{equation}
to find the ${\rm Re}\,\sigma^{reg} (\omega)$ weight distribution
in the vicinity of the upper Hubbard band lower-limit point
located at ($k=0,\,\omega=Eu=E_{MH}$). Use of this relation
reveals that the exponent (\ref{zetadsf}) controls the frequency
dependence of that weight distribution. Combination of Eqs.
(\ref{Ichiunrho}) and (\ref{condDSSF}) for $n\rightarrow 1$ and
$E_u\rightarrow E_{MH}$ leads to,

\begin{equation}
{\rm Re}\,\sigma^{reg} (\omega)\propto (\omega - E_{MH})^{1/2} \,
. \label{opcondn1}
\end{equation}
This expression applies to all finite values of $U/t$.

In contrast to the upper Hubbard band dynamical structure factor
expression (\ref{Ichiunrho}) which refers to $k=\pi\mp 2k_F$ and
values of $\omega$ such that $(\omega - E_u)$ is small, the
frequency dependence (\ref{opcondn1}) of the $k=0$ optical
conductivity in the vicinity of the upper Hubbard band lower-limit
point was studied by other methods \cite{optical,opticalthey}. In
reference \cite{optical} that zero-momentum conductivity edge was
studied for the metallic phase, whose critical exponent is
different from the exponent (\ref{zetadsf}). However, in the limit
$n\rightarrow 1$ the zero-momentum density dependent exponent
found in such a reference provides the correct half-filling
exponent. In spite of the different method used in the evaluation
of that exponent, the half-filling expression derived in Ref.
\cite{optical} coincides with that of Eq. (\ref{opcondn1}).
Moreover, the same optical conductivity edge was studied in Ref.
\cite{opticalthey} for both small and large values of $U/t$. These
investigations also lead to the same critical exponent $1/2$ for
both these limits, in agreement with expression (\ref{opcondn1})
which is valid for all finite values of $U/t$. As the conductivity
exponent $-1/2$ is characteristic of a finite-energy semiconductor
edge, also the conductivity exponent $1/2$ is now believed to be
an universal signature of the Mott-Hubbard insulator
\cite{optical,opticalthey}.

\subsection{SPIN-SINGLET COOPER-PAIR FIRST AND SECOND UPPER-HUBBARD BAND WEIGHT DISTRIBUTIONS}

The energy $E_u$ becomes small for low values of $U/t$ and
electronic densities $n$ in the vicinity of one. Thus, it is
interesting to clarify whether there are singular spectral
features in the Cooper-pair spectral function at the upper-Hubbard
bands lower limit. Indeed, for the above values of $U/t$ and
electronic density such singular features could lead to a
superconductivity instability for a system of weakly coupled
Hubbard chains. Unfortunately, the only singular feature found
below is a $\delta$ function peak located at the point
$(k=\pi,\,\omega =E_u)$. That isolated peak results from the
$\eta$-pairing mechanism \cite{HL}. As the electronic density $n$
approaches one the weight of that peak vanishes as $(1-n)$. While
this spectral structure cannot lead to a superconductivity
instability for the coupled-chain system, other singular features
for excitation energy above the first upper Hubbard band lower
limit could exist, as further discussed in Sec. V.

According to Eqs. (\ref{DJcDJs}) and Eqs. (51) and (52) of Ref.
\cite{IIa} with $\Delta L_{s,\,-1/2}=0$ and $\Delta N_{s\nu} =0$
for $\nu >1$, in the case of the $M_{c,\,-1/2}=M=1$ type of
transitions there is for the spin-singlet Cooper-pair spectral
function a one-$-1/2$ Yang holon type of transition and three
one-$q= \pm q^0_{c1} = \pm[\pi -2k_F]$ $c1$ pseudoparticle types
of transitions. However, we find that the one-$q= \pm q^0_{c1} =
\pm[\pi -2k_F]$ $c1$ pseudoparticle transition whose excitation
momentum and energy is the same as the momentum and energy of the
one-$-1/2$ Yang holon type of transition does not contribute to
the singlet Cooper pair spectral function. Thus, only the
one-$-1/2$ Yang holon type of transition and two of the three
one-$q= \pm q^0_{c1} = \pm[\pi -2k_F]$ $c1$ pseudoparticle types
of transitions contribute to weight-distribution features. We find
that the weight-distribution feature generated by the one-$-1/2$
Yang holon transition is a single $\delta$ peak located in the
first upper Hubbard band lower limit at $k=\pi$ and $\omega=E_u$.
The matrix element between the corresponding $c1$ pseudoparticle
excited state with the same momentum $k=\pi$ and excitation energy
$\omega=E_u$ and the ground state vanishes in this case.
Therefore, this singlet Cooper pair spectral function feature
results from the one-$-1/2$ Yang holon excited state only. There
are also two first upper Hubbard band lower-limit points located
at ($k=\pi \mp 4k_F,\,\omega =E_u$) which are generated by one-$q=
\pm q^0_{c1} = \pm[\pi -2k_F]$ $c1$ pseudoparticle transitions.
Below we also study the weight distribution in the vicinity of
these first Hubbard band lower-limit points.

For the one-$-1/2$ Yang holon transition we find that $\Delta N_c
=\Delta N_s = \Delta J_c = \Delta J_s=0$. Thus, this is an example
of a transition where the quantity (\ref{zeta*}) reads
$\zeta_{ss}=-2$ and is not a critical exponent. Instead, the
correlation function at that point of the $(k,\,\omega)$-plane is
of the form given in Eq. (\ref{Ichiun-1-2}) for all values of
$U/t$ and reads,

\begin{equation}
\chi_{ss}(\pi,\,\omega) = {C_{\rho}\over \omega - E_u - i\beta} \,
, \label{chiss}
\end{equation}
where $\beta\rightarrow 0$ and $C_{\rho}\geq 0$ is a constant. It
follows that the associated singlet Cooper pair spectral-function
weight distribution is indeed of the type given in Eq.
(\ref{Ichiun-1-2}) and reads,

\begin{equation}
{\rm Im}\chi_{ss }(\pi,\omega) = \pi\,C_{\rho}\,\delta (\omega -
E_u) \, . \label{Imchiss}
\end{equation}

The weight feature (\ref{Imchiss}) can be obtained by direct
evaluation of the matrix elements on the right-hand side of
expression (\ref{corrpos}) for $\vartheta =ss$. That procedure
confirms in this particular case the validity of the results
obtained by our method. Moreover, direct calculation of the matrix
elements reveals that for $k=\pi$ the expressions (\ref{chiss})
and (\ref{Imchiss}) do not refer to values of $\omega$ such that
$(\omega -E_u)$ is small only, but to {\it all} values of
excitation energy $\omega$. Direct evaluation of the matrix
elements also reveals that the constant $C_{\rho}$ of Eqs.
(\ref{chiss}) and (\ref{Imchiss}) reads $C_{\rho}=[N_a-N]$ and
vanishes for the Mott-Hubbard insulator.

In order to evaluate the matrix elements of expression
(\ref{corrpos}) for $\vartheta =ss$ and in the particular case of
the momentum value $k=\pi$, we note that the off diagonal
generator ${\hat{S}}^c_{+}$ of the $\eta$-spin $SU(2)$ algebra
given in Eq. (9) of Ref. \cite{I} can be rewritten as follows,

\begin{equation}
{\hat{S}}^c_{+}= \sum_{k'}
c^{\dagger}_{k',\,\downarrow}\,c^{\dagger}_{\pi-k',\,\uparrow} \,
. \label{Sckk'}
\end{equation}
Application of that generator onto a ground state produces an
excited state with a $-1/2$ Yang holon. The generator of the above
transition corresponds to creation of that $-1/2$ Yang holon and
involves the initial ground state and this single final excited
state only. After normalization this excited state can be written
as follows,

\begin{equation}
\vert\,L_{c,\,-1/2}=1\rangle = {\sum_{k'}
c^{\dagger}_{k',\,\downarrow}\,c^{\dagger}_{\pi-k',\,\uparrow}
\over\sqrt{N_a-N}}\,\vert\,GS\rangle \, . \label{exL}
\end{equation}

The main point is that at $k=\pi$ the singlet superconductivity
operator given in Eq. (\ref{Ok}) reads,

\begin{equation}
{\hat{\cal{O}}}_{ss} (\pi) = {\hat{S}}^c_{+}= \sum_{k'}
c^{\dagger}_{k',\,\downarrow}\,c^{\dagger}_{\pi-k',\,\uparrow} \,
. \label{OrhoSckk'}
\end{equation}
Use of Eq. (\ref{OrhoSckk'}) in Eq. (\ref{corrpos}) reveals that
for $k=\pi$ the correlation function expression results from the
overlap of the ground state with the excited state (\ref{exL})
only. Thus, for $k=\pi$ only one matrix element between the ground
state and the available excited states of momentum $k=\pi$ is
finite. That matrix element corresponds to the state (\ref{exL}).
It follows that for $k=\pi$ the singlet superconductivity
correlation function and corresponding spectral function are
indeed given by expressions (\ref{chiss}) and (\ref{Imchiss}) with
$C_{\rho}=[N_a-N]$ for all values of $\omega$.

The one-$-1/2$ Yang holon state (\ref{exL}) is associated with the
$\eta$-pairing mechanism \cite{HL}. That mechanism leads in the
metallic phase to the $\delta$-function peak (\ref{Imchiss}) in
the singlet Cooper pair spectral function. Such a state possesses
off-diagonal long-range order \cite{HL}. The corresponding
$\delta$ peak is located at finite excitation energy. The $\delta$
peak (\ref{Imchiss}) is absent in the case of the Mott-Hubbard
insulator where $C_{\rho}=[N_a-N]\rightarrow 0$. This vanishing is
required by a half-filling selection rule, since the ground-state
$\eta$ spin value of the Mott-Hubbard insulator reads $S_c=0$.
Note that the present transition is a mere rotation in $\eta$-spin
space which conserves $S_c$ and leads to $\Delta S_c=0$ and
$\Delta S_c^z=+1$. Thus, only for initial ground states such that
$S_c\geq 1$ can this transition occur. This excludes both the
Mott-Hubbard insulator ground state such that $S_c=0$ and the
one-hole doped Mott-Hubbard insulator ground state such that
$S_c=1/2$.

Next, we consider the weight distribution in the vicinity of the
first upper Hubbard band lower-limit points generated by other
$M_{c,\,-1/2}=M=1$ types of transitions. These transitions
correspond to a region of little spectral weight and are such that
$N_{c1}=1$, $\Delta J_c=\pm 1/2$, $J_{c1}=\mp 1/2$, and $\Delta
N_c =\Delta N_s =\Delta J_s=0$. Use of the general expression
(\ref{Ichiun}) leads in the limit $m\rightarrow 0$ to the
following expression for the spin singlet Cooper pair spectral
function for momentum values $k=\pi \mp 4k_F$ and excitation
energy $\omega$ such that $(\omega -E_u)$ is small,

\begin{equation}
{\rm Im}\,\chi_{ss} (\pi\mp 4k_F,\,\omega)\propto (\omega -
E_u)^{\zeta_{ss}} \, . \label{Ichiunss2}
\end{equation}
This weight-distribution edge does not occur for the Mott-Hubbard
insulator. From use of Eq. (\ref{zeta*}) we find the following
expression for the critical exponent $\zeta_{ss}$ in the limit
$m\rightarrow 0$,

\begin{equation}
\zeta_{ss} = -2\Bigl[1-[\sqrt{2K_{\rho}}]^2\Bigr]  \, .
\label{zetass}
\end{equation}
This exponent is a function of $U/t$ which for all electronic
densities changes from $\zeta_{ss}\rightarrow 2$ as
$U/t\rightarrow 0$ to $\zeta_{ss}\rightarrow 0$ as
$U/t\rightarrow\infty$. Such a dependence on $U/t$ is continuous.
(Since there is not much spectral weight in the vicinity of this
edge feature, we do not plot the positive exponent
(\ref{zetass}).)

We close our study by considering the weight distribution in the
vicinity of the $M_{c,-1/2}=M=2$ second upper Hubbard band
lower-limit points located at excitation energy $\omega=2E_u$. We
note that for the Mott-Hubbard insulator there is no first upper
Hubbard band. In this case there is an energy gap which equals
twice the Mott-Hubbard gap and separates the singlet Cooper pair
removal and addition spectral functions. The latter function
corresponds to the present second upper Hubbard band in the limit
$n\rightarrow 1$.

The $M_{c,\,-1/2}=M=2$ types of transitions are such that $\Delta
N_c=-2$, $\Delta N_s=-1$, and $\Delta J_s=0$. Interestingly, there
are five different types of transitions which contribute to the
same weight distribution in the vicinity of two points and lead to
the same critical exponent. These five types of transitions are
such that (a) $L_{c,\,-1/2}=2$ and $\Delta J_c=\pm 1/2$, (b)
$L_{c,\,-1/2}=1$, $N_{c1}=1$, $\Delta J_c=0$, $J_{c1}=\pm 1/2$,
(c) $N_{c1}=2$, $\Delta J_c=\pm 1/2$, $J_{c1}=0$, (d) $N_{c1}=2$,
$\Delta J_c=\pm 1/2$, $J_{c1}=\pm 1$, and (e) $N_{c,\,2}=1$,
$\Delta J_c=0$, $J_{c1}=\pm 1/2$. All transitions of these types
contribute to the weight distribution around the two second upper
Hubbard band lower-limit points located at ($k=\pm 2k_F,\,\omega
=2E_u$). In addition, there is a sixth type of transition which
leads to two second upper Hubbard band lower-limit points located
at ($k=\pm 6k_F,\,\omega =2E_u$). This type of transition is such
that $N_{c1}=2$, $\Delta J_c=\pm 1/2$, and $J_{c1}=\mp 1/2$. There
is nearly no spectral weight in the vicinity of these two latter
points.

Use of the general expression (\ref{Ichiun}) leads in the limit
$m\rightarrow 0$ to the following expression for the spin singlet
Cooper pair spectral function for excitation energy $\omega$ such
that $(\omega -2E_u)$ is small and momentum values $k=k_2^l =
l\,2k_F$ where $l=\pm 1,\,\pm 3$,

\begin{equation}
{\rm Im}\,\chi_{ss} (l\,2k_F,\,\omega)\propto (\omega -
2E_u)^{\zeta_{ss}^{\vert l\vert}} \, ; \hspace{0.5cm} l=\pm
1,\,\pm 3 \, . \label{Ichiunss3}
\end{equation}
>From use of Eq. (\ref{zeta*}), we find the following expression
for the critical exponent $\zeta_{ss}^{\vert l\vert}$ in the limit
$m\rightarrow 0$,

\begin{equation}
\zeta_{ss}^{\vert l\vert} = -2\Bigl[1 - [l\,\sqrt{2K_{\rho}}/2]^2
- [1/\sqrt{2K_{\rho}}]^2\Bigr]  \, ; \hspace{0.5cm} l=\pm 1,\,\pm
3 \, . \label{zetadss3}
\end{equation}
Note that for $l=\pm 1$ this exponent equals the dynamical
structure factor exponent $\zeta_{\rho}$ given in Eq.
(\ref{zetadsf}). Thus, it changes from $\zeta_{ss}^1\rightarrow 0$
as $U/t\rightarrow 0$ to $\zeta_{ss}^1\rightarrow 1/2$ as
$U/t\rightarrow\infty$. It is plotted in Fig. 3 as a function of
the electronic density $n$ for different values of $U/t$. On the
other hand, the exponent $\zeta_{ss}^3$ is much larger and changes
from $\zeta_{ss}^3\rightarrow 8$ as $U/t\rightarrow 0$ to
$\zeta_{ss}^1\rightarrow 9/2$ as $U/t\rightarrow\infty$. This
exponent corresponds to a region of very little spectral weight.
In the limit $n\rightarrow 1$ the four second upper Hubbard
lower-limit points located at ($k=\pm 2k_F,\,\omega =2E_u$) and
($k=\pm 6k_F,\,\omega =2E_u$) become the same single point. Thus,
in this case the smaller exponent $\zeta_{ss}^1$ corresponds to
the dominant contribution and controls the weight distribution in
the vicinity of the lower-limit point located at ($k=\pi,\,\omega
=2Eu=2E_{MH}$). For the $n=1$ Mott-Hubbard insulator phase the
exponent $\zeta_{ss}^1$ reads $1/2$ for all finite values of
$U/t$. It equals the exponent which controls the weight
distribution around the zero-energy point located at
($k=\pi,\,\omega =0$). Interestingly, comparison with the results
of Ref. \cite{Carmelo97pp} reveals that also in the metallic phase
the exponents (\ref{zetadss3}) for $l=\pm 1$ and $l=\pm 3$ equal
the corresponding exponent that control the low-energy weight
distribution of the singlet Cooper pair removal spectral function
in the vicinity of the zero-energy points located at ($k=\pm
2k_F,\,\omega =0$) and ($k=\pm 6k_F,\,\omega =0$), respectively.
However, in the metallic phase there is more spectral weight in
the vicinity of the latter zero-energy points than in the vicinity
of the corresponding second upper Hubbard band lower-limit points.

Finally, we note that if one extends the present study to
finite-energy excited states with finite occupancies of $-1/2$ HL
spinons and/or $s\nu$ pseudoparticles belonging to $\nu>1$
branches, similar features are obtained for gapped spin
excitations. For instance, in the case of the spin-down triplet
superconductivity spectral function a $\delta$-function peak
similar to (\ref{Imchiss}) is obtained at $k=\pi$ and $\omega=E_u
+\mu_0\,H$. This peak is generated by a transition from the ground
state to a single finite-energy excited state with one $-1/2$ Yang
holon and one $-1/2$ HL spinon.

\section{CONCLUDING REMARKS}

In this paper we derived general few-electron spectral function
expressions for the 1D Hubbard model in the vicinity of the lower
limit of the upper Hubbard bands. For the one electron addition
and dynamical structure factor we studied the weight distribution
in the vicinity of the lower limit of the first upper Hubbard
band. In the case of the singlet Cooper pair spectral function we
considered the same problem in the vicinity of the lower limits of
both the first and second upper bands. The weight of these Hubbard
bands is generated by dominant holon and spinon processes which
amount to more than 99\% of the spectral weight \cite{V}. Our
general expressions were obtained by combination of symmetries of
the 1D Hubbard model, such as the holon and spinon numbers
conservation laws, with the finite-energy holon and spinon
description of the quantum problem recently introduced in Refs.
\cite{I,IIIb}.

The results obtained in this paper provide physically interesting
and useful information about the finite-energy spectral properties
of the present many-electron 1D quantum liquid and are
complementary to the branch line studies of Refs.
\cite{spectral0,spectral,V}. In the case of the one-electron
problem the combination of our results with those of Ref.
\cite{spectral} provides all finite-energy spectral-weight
singularities. Since the singular spectral features observed in
quasi-1D metals agree quantitatively with the model singular
branch lines, our results are of interest for the further
understanding of the unusual spectral properties observed in
low-dimensional materials
\cite{Menzel,Hasan,Ralph,spectral0,spectral}. Moreover, from the
finite-energy weight-distribution found for the dynamical
structure factor, we checked that in the limit of half filling our
results lead for all finite values of $U/t$ to the expected
Mott-Hubbard insulator exponent $1/2$ for the onset of the
finite-frequency optical conductivity absorption. Unfortunately,
we found no singular features at the the upper-Hubbard bands lower
limit of the spin-singlet Copper-pair spectral function other than
the expected $\eta$-pairing $\delta$-function peak. However, it
could be that there are branch lines just above such a lower limit
which show singular spectral features. Since the energy width of
the first Hubbard band is small for low values of $U/t$ and
electronic densities $n$ in the vicinity of one, such singular
features could lead to an instability for a system of weakly
coupled Hubbard chains. We thus suggest that this problem is
further studied by the branch-line method used in Ref.
\cite{spectral} for the one-electron problem.


\ack
We thank Jo\~ao Lopes dos Santos and Karlo Penc for stimulating
discussions. In its initial stage this research was supported by
PRAXIS under Grant No. BD/3797/94.

\appendix

\section{THE FINITE-SIZE ENERGY SPECTRUM QUANTITIES}

In this Appendix we show that the functional
$2\,\Delta^{\iota}_{\alpha}$ has the form given in Eq. (\ref{cd2})
and discuss issues related to the vanishing of the $c\nu$
finite-size energy spectrum for $q=\pm q^0_{c\nu} = \pm[\pi
-2k_F]$.

By direct use of the results of Ref. \cite{IIIb}, we find that for
the excited states that span the reduced J-CPHS ensemble subspaces
${\cal H}_{red}$ the functional (\ref{Dia}) can be written as,
\bea
2\,\Delta^{\iota}_{\alpha} &=& \Bigl[\iota\sum_{\alpha'=c,s}\,
\xi^0_{\alpha,\,\alpha'}\,{\Delta N_{\alpha'}\over 2} +
\iota\sum_{\nu=1}^{\infty}\,\xi^0_{\alpha,\,c\nu}\,{N_{c\nu}\over
2} \nonumber \\
&+& \,\sum_{\alpha'=c,s}\, \xi^1_{\alpha,\,\alpha'}\,\Delta
J_{\alpha'} - \sum_{\nu=1}^{\infty}\,\xi^1_{\alpha,\,c\nu}\,
J_{c\nu}\Bigr]^2 \, . \label{cd}
\eea
Here the parameters $\xi^j_{\alpha,\,\alpha'}$ and
$\xi^j_{\alpha,\,c\nu}$ can expressed in terms of the
two-pseudofermion phase shifts defined in Ref. \cite{IIIb} as
follows,
\bea
\xi^j_{\alpha,\,\alpha'} &=& \delta_{\alpha,\,\alpha'} + \sum_{l=\pm
1}(l^j)\,\Phi_{\alpha,\,\alpha'} (q^0_{F\alpha},\,l\,
q^0_{F\alpha'}) \, \nonumber \\
\xi^j_{\alpha,\,c\nu} &=&
\sum_{l=\pm 1}(l^j)\,\Phi_{\alpha,\,c\nu} (q^0_{F\alpha},\,l\,[\pi
-2k_F]) \, . \label{xiGS-xianu}
\eea
In the limit $m\rightarrow 0$ these parameters read $\xi^0_{c,\,c}
= 1/\sqrt{2K_{\rho}}$, $\xi^0_{c,\,s}=0$,
$\xi^0_{s,\,c}=-1/\sqrt{2}$, $\xi^0_{s,\,s}=\sqrt{2}$,
$\xi^0_{c,\,c\nu}=0$, $\xi^0_{s,\,c\nu}=0$, $\xi^1_{c,\,c}  =
\sqrt{2K_{\rho}}$, $\xi^1_{c,\,s} = \sqrt{K_{\rho}/2}$, $
\xi^1_{s,\,c}=0$, $\xi^1_{s,\,s}=1/\sqrt{2}$,
$\xi^1_{c,\,c\nu}=\sqrt{2K_{\rho}}$, and $\xi^1_{s,\,c\nu}=0$.
Here $K_{\rho}$ is the parameter defined in Ref. \cite{Schulz}
which is such that $K_{\rho}\rightarrow 1$ as $U/t\rightarrow 0$
and $K_{\rho}\rightarrow 1/2$ as $U/t\rightarrow\infty$.

The creation of $-1/2$ Yang holons does not lead to any
contribution to the value of the functional (\ref{cd}). However,
the creation of $c\nu$ pseudoparticles leads to contributions
through the terms
$\iota\sum_{\nu=1}^{\infty}\,\xi^0_{\alpha,\,c\nu}\,N_{c\nu}/2$
and $-\sum_{\nu=1}^{\infty}\,\xi^1_{\alpha,\,c\nu}\, J_{c\nu}$ of
that functional. On the other hand, the symmetry discussed in the
text below Eq. (\ref{cd2}) requires that the values of the
phase-shift parameters $\xi^j_{\alpha,\,c\nu}$ of Eq.
(\ref{xiGS-xianu}) are such that the functional (\ref{cd})
simplifies to (\ref{cd2}). Indeed, by manipulation of the integral
equations of Ref. \cite{IIIb} that define the two-pseudofermion
phase shifts, we find the following relation between the
parameters $\xi^j_{\alpha,\,c\nu}$ and $\xi^1_{\alpha,\,c}$
defined in Eq. (\ref{xiGS-xianu}),

\begin{equation}
\xi^j_{\alpha,\,c\nu}= j\,\xi^1_{\alpha,\,c} \, ; \hspace{0.5cm}
\alpha=c,\,s \, ; \hspace{0.3cm} \nu=1,\,2,\,3,... \, ;
\hspace{0.3cm} j=0,\,1 \, . \label{xianuEX}
\end{equation}
Use of this relation in Eq. (\ref{cd}) leads to (\ref{cd2}).

Moreover, the same symmetry is behind the vanishing of the
finite-size energy contributions of the $\pm q^0_{c\nu} = \pm[\pi
-2k_F]$ $c\nu$ pseudoparticles. Indeed, by careful analysis of the
general energy spectrum $\Delta E_{GL}$ we find that these
contributions vanish provided that the quantity,

\begin{equation}
2\,\Delta^{\iota}_{c\nu} = \Bigl[{N_a\over 2\pi}\,\Bigl(
\iota\,\Delta q^0_{c\nu}+ {Q_{c\nu} (\iota\,q^0_{c\nu})\over
N_a}\Bigr)\Bigr]^2 \, ; \hspace{0.5cm} \nu=1,2,...\, ;
\hspace{0.3cm} \iota = \pm 1 \, , \label{Dicnu}
\end{equation}
also vanishes. Here $Q_{c\nu} (q)$ is the functional defined by
Eq. (73) of Ref. \cite{IIIb}. Use of the equality $Q_{c\nu}
(\iota\,q^0_{c\nu})/N_a=-\iota\,\Delta q^0_{c\nu}$ given in Eq.
(126) of the same reference in expression (\ref{Dicnu}) confirms
that $2\,\Delta^{\iota}_{c\nu}$ vanishes for the excited states
that span the reduced subspace.

These properties apply only to $c\nu$ pseudoparticles such that
$q\rightarrow \pm[\pi -2k_F]$. Indeed, these quantum objects
become non-interacting and localized in that limit. This behavior
is related to the fact the $-1/2$ Yang holon is also
non-interacting and invariant under the electron -
rotated-electron unitary transformation \cite{IIa}. However, in
general a $c\nu$ pseudoparticle is different from the rotated
$c\nu$ pseudoparticle. The only exception is precisely for bare
momentum values $q$ such that $q\rightarrow \pm q^0_{c\nu} =
\pm[\pi -2k_F]$. The $c\nu$ pseudoparticles become non-interacting
in that limit only. Indeed, the energy associated with creation of
a $-1/2$ Yang holon is given by $E_u$. A $c\nu$ pseudoparticle is
a composite quantum object of $\nu$ $-1/2$ holons and $\nu$ $+1/2$
holons. Creation of a $+1/2$ holon is a process which requires no
energy. The amount of energy required for creation of a $c\nu$
pseudoparticle is $\nu\,E_u +\epsilon^0_{c\nu}(q)$. Thus, the
energy per $-1/2$ holon is $E_u +\epsilon^0_{c\nu}(q)/\nu\neq
E_u$. However, one has that $\epsilon^0_{c\nu}(q)\rightarrow 0$ as
$q\rightarrow \pm q^0_{c\nu} = \pm[\pi -2k_F]$ and thus in this
limit the energy per $-1/2$ holon becomes $E_u$ and equals that of
a non-interacting $-1/2$ Yang holon.

\section{THE FINITE-ENERGY CORRELATION FUNCTION}

Here we show that the asymptotic expansion of the correlation
function $\langle
GS\vert\,{\hat{\phi}}_{\vartheta}(x,\,t)\,{\hat{\phi}}_{\vartheta}(0,0)\,\vert\,GS\rangle$
is of the form given in Eq. (\ref{correl}). The operator
${\hat{\phi}}^{GL}_{\vartheta}(x,\,t)$ acts and is defined in a
reduced J-CPHS ensemble subspace associated with small values of
the excitation energy $\Delta E_{GL}=(\omega-\omega_{HS})$ and
excitation momentum $\Delta P_{GL}=(k-k_M^l)$. In the Heisenberg
description and in that reduced Hilbert subspace the space
coordinate $x$ and time $t$ dependent physical field
${\hat{\phi}}^{GL}_{\vartheta}(x,\,t)$ can be expressed relative
to its $x=0$ and $t=0$ expression
${\hat{\phi}}_{\vartheta}(0,\,0)$ as,

\begin{equation}
{\hat{\phi}}^{GL}_{\vartheta}(x,\,t)=e^{-i[:\hat{H}_{GL}:\,t
-:\hat{P}_{GL}:\,x]}\,{\hat{\phi}}_{\vartheta}(0,\,0)\,e^{i[:\hat{H}_{GL}:\,t
-:\hat{P}_{GL}:\,x]} \, . \label{tconf}
\end{equation}
For the Hamiltonian $:{\hat{H}}_{GL}:$ of Eq. (\ref{Hno2-nhHGL})
the low-energy energy spectrum (\ref{DEGLCF}) and momentum
spectrum $\Delta P_{GL} = {2\pi\over N_a}\,\sum_{\alpha
=c,s}\,\sum_{\iota =\pm 1}\,\iota\,[\,\Delta^{\iota}_{\alpha} +
N^{ph}_{\alpha,\,\iota}]$ are conformal invariant and correspond
to a two-component conformal-field theory associated with the $c$
and $s$ pseudoparticle excitation branches
\cite{Belavin,Carmelo97pp}. The effect of the $c\nu$
pseudoparticles is merely to shift the $c$ pseudoparticle current
number deviations $\Delta J_c$ by
$-\sum_{\nu=1}^{\infty}J_{c\nu}$. Fortunately, such an effect does
not affect the conformal invariance of the low-energy energy
spectrum and momentum spectrum. Thus, the asymptotic expression
for the low-energy correlation function of the physical field
${\hat{\phi}}^{GL}_{\vartheta}(x,\,t)$ is for electronic and spin
densities such that $0<n<1$ and $0<m<n$, respectively, of the
following general form \cite{Belavin,Carmelo97pp},

\begin{equation}
\langle GS\vert {\hat{\phi}}_{\vartheta}^{GL}(x,\,t)\,
{\hat{\phi}}_{\vartheta}^{GL}(0,\,0)\vert GS\rangle\propto
\prod_{\alpha =c,\,s}\,\prod_{\iota=\pm 1} {1\over (x-\iota
v_{\alpha}\,t)^{2\Delta^{\iota}_{\alpha}}} \,  . \label{cf}
\end{equation}

We recall that the 1D Hubbard model $:\hat{H}:$ and Hamiltonian
$:{\hat{H}}_{GL}:$ of Eq. (\ref{Hno2-nhHGL}) refer to the same
ground state. Thus, the field ${\hat{\phi}}_{\vartheta}(x,\,t)$ is
such that,

\begin{equation}
{\hat{\phi}}_{\vartheta}(x,\,t)=e^{-i[:\hat{H}:\,t-:\hat{P}:\,x]}\,
{\hat{\phi}}_{\vartheta}(0,\,0)\,e^{i[:\hat{H}:\,t-:\hat{P}:\,x]}
\, , \label{t}
\end{equation}
where $:\hat{P}:$ is the momentum operator of Eq.
(\ref{DPop-k0op}). Since the Hamiltonians $:\hat{H}:$ and
$:{\hat{H}}_{GL}:$ of Eq. (\ref{Hno2-nhHGL}) have the same energy
eigenstates, there is a one-to-one correspondence between the
correlation function terms (\ref{cf}) of the Hamiltonian
$:{\hat{H}}_{GL}:$ and those of the 1D Hubbard model $:\hat{H}:$.
Transitions to different reduced subspaces ${\cal{H}}_{red}$ lead
to different correlation-function terms. Moreover, transitions to
reduced subspaces with different values for $M\,E_u$ and $k_M^l$
lead for the 1D Hubbard model to correlation-function
contributions with different values of the excitation energy
$\omega\approx M\,E_u$ and momentum $k\approx k_M^l$. Thus, our
first choice is the energy value $M\,E_u$ and momentum value
$k_M^l$ which our correlation-function asymptotic term refers to.
If for the same values of $M\,E_u$ and $k_M^l$ there are several
reduced subspaces, we choose that associated with the
leading-order asymptotic correlation-function term for the
Hamiltonian $:{\hat{H}}_{GL}:$ of Eq. (\ref{Hno2-nhHGL}). That
leading-order term is of the form (\ref{cf}). Let $\vert\,\psi
(j)\rangle$ with $j=1,2,...$ denote the excited states that span
the chosen reduced Hilbert subspace ${\cal{H}}_{red}$. Based on
the commutation relations of the three normal-ordered Hamiltonians
related by Eq. (\ref{Hno2-nhHGL}) and three normal-ordered
momentum operators involved in Eq. (\ref{DPop-k0op}) we find that,

\begin{eqnarray}
{\hat{\phi}}_{\vartheta}(x,\,t) & = &
e^{-i[({\hat{H}}_{HS}+:{\hat{H}}_{GL}:)\,t-
({\hat{P}}_0+:{\hat{P}}_{GL}:)\,x]}\,
{\hat{\phi}}_{\vartheta}(0,\,0)\,e^{i[({\hat{H}}_{HS}+:{\hat{H}}_{GL}:)\,t
-({\hat{P}}_0+:{\hat{P}}_{GL}:)\,x]}
\nonumber\\
& = & e^{-i[{\hat{H}}_{HS}\,t-{\hat{P}}_0\,x]}\,
e^{-i[{:\hat{H}}_{GL}:\,t-:{\hat{P}}_{GL}:\,x]}
{\hat{\phi}}_{\vartheta}(0,0)e^{i[{\hat{H}}_{GL}\,t-{\hat{P}}_{GL}\,x]}\,
e^{i[{\hat{H}}_{HS}\,t-{\hat{P}}_0\,x]}
\nonumber\\
& = & e^{-i[{\hat{H}}_{HS}\,t-{\hat{P}}_0\,x]}\,
 {\hat{\phi}}_{\vartheta}^{GL}(x,\,t) e^{i[{\hat{H}}_{HS}\,t-{\hat{P}}_0\,x]}
\, , \label{pppp}
\end{eqnarray}
where we have used Eqs. (\ref{tconf}) and (\ref{t}). Equation
(\ref{pppp}) confirms that ${\hat{\phi}}_{\vartheta}(0,\,0) =
{\hat{\phi}}_{\vartheta}^{GL}(0,\,0)$.

Once the ground state has eigenvalue zero both for
${\hat{H}}_{HS}$ and ${\hat{P}}_0$ and all excited states that
span the subspace ${\cal{H}}_{red}$ have for these operators {\it
the same} eigenvalues $M\,E_u$ and $k_M^l$, respectively, we find
that,

\begin{eqnarray}
& &
\langle
GS\vert{\hat{\phi}}_{\vartheta}(x,\,t)\,{\hat{\phi}}_{\vartheta}(0,\,0)\vert
GS\rangle  \nonumber \\
& = & \langle GS\vert\,
e^{-i[{\hat{H}}_{HS}\,t-{\hat{P}}_0\,x]}\,
{\hat{\phi}}_{\vartheta}^{GL}(x,t)\,
e^{i[{\hat{H}}_{HS}\,t-{\hat{P}}_0\,x]}{\hat{\phi}}_{\vartheta}^{GL}(0,\,0)\vert
GS \rangle \nonumber \\
& = & \sum_{j} \langle GS\vert
{\hat{\phi}}_{\vartheta}^{GL}(x,\,t)\,
e^{i[{\hat{H}}_{HS}\,t-{\hat{P}}_0\,x]}\,\vert\psi (j)\rangle
\langle\psi (j)\vert{\hat{\phi}}_{\vartheta}^{GL}(0,\,0)\vert GS
\rangle
\nonumber \\
& = & e^{i[M\,E_u\,t-k_M^l\,x]}\,\langle
GS\vert{\hat{\phi}}_{\vartheta}^{GL}(x,\,t)\,
{\hat{\phi}}_{\vartheta}^{GL}(0,\,0)\vert GS\rangle \, .
\label{relation}
\end{eqnarray}
Here the $j$ summation refers to the above set of excited states
$\{\vert\psi (j)\rangle \}$ which span the suitable reduced
subspace ${\cal{H}}_{red}$. Finally, the combination of Eqs.
(\ref{cf}) and (\ref{relation}) implies that the leading term in
the asymptotic expansion of the correlation function of the 1D
Hubbard model is of the form given in Eq. (\ref{correl}).


\end{document}